\documentclass[epj]{svjour}
\usepackage{epsfig}
\usepackage{amssymb}
\usepackage{amsmath, bbm}
\usepackage[figuresright]{rotating}
\usepackage{capt-of}
\usepackage[numbers,sort&compress]{natbib}

\newlength{\feynwidth} 

\newcommand{\La}{{\Lambda}}
\newcommand{\Si}{{\Sigma}}

\newcommand{\be}{\begin{eqnarray}}
\newcommand{\ee}{\end{eqnarray}}

\usepackage{color}
\usepackage{hyperref}

\begin{document}

\title{Hyperon-nucleon interaction within chiral effective field theory revisited}
\titlerunning{Hyperon-nucleon interaction}

\author{J. Haidenbauer$^{1}$, U.-G. Mei{\ss}ner$^{2,1,3,4}$, A. Nogga$^{1,4}$}
\authorrunning{J. Haidenbauer et al.}

\institute{
$^1$Institute for Advanced Simulation,
Institut f\"ur Kernphysik (Theorie) and J\"ulich Center for
Hadron Physics, Forschungszentrum J\"ulich, D-52425 J\"ulich, Germany \\
$^2$Helmholtz Institut f\"ur Strahlen- und Kernphysik and Bethe Center
for Theoretical Physics, Universit\"at Bonn, D-53115 Bonn, Germany\\
$^3$Tbilisi State  University,  0186 Tbilisi, Georgia\\
$^4$JARA-High Performance Computing, Forschungszentrum J\"ulich, 
D-52425 J\"ulich, Germany
}
\date{Received: date / Revised version: date}
\abstract{The $\Lambda N$ and $\Sigma N$ interactions
are considered at next-to-leading order in SU(3) chiral 
effective field theory. Different options for the 
low-energy constants that determine the strength of the 
contact interactions are explored. Two variants are
analysed in detail which yield equivalent results for 
$\Lambda N$ and $\Sigma N$ scattering observables but
differ in the strength of the $\Lambda N \to \Sigma N$
transition potential. The influence of this difference
on predictions for light hypernuclei and on the properties
of the $\Lambda$ and $\Sigma$ hyperons in nuclear matter 
is investigated and discussed. The effect of the variation 
in the potential strength of the $\Lambda N$-$\Sigma N$ 
coupling (also called $\Lambda -\Sigma$ conversion) is found
to be moderate for the considered $^3_\Lambda \rm H$ and 
$^4_\Lambda \rm He$ hypernuclei but 
sizable in case of the matter properties. Further, the size of  
three-body forces  and their relation to different 
approaches to hypernuclear interactions is discussed. 
\PACS{
      {12.39.Fe}{Chiral Lagrangians}   \and
      {13.75.Ev}{Hyperon-nucleon interactions}   \and
      {21.30.Fe}{Forces in hadronic systems and effective interactions}
     }
}

\maketitle

\section{Introduction} 
\label{sec:1}

In 2013 the J\"ulich-Bonn-Munich group presented a study of $\Lambda N$ 
and $\Sigma N$ scattering up to next-to-leading order (NLO) in SU(3) chiral 
effective field theory (EFT)~\cite{Haidenbauer:2013}, following 
closely earlier analogous investigations of the $NN$ interaction
\cite{Epelbaum:2005,Epelbaum:2006,Epelbaum:2008,Machleidt:2011}. 
It demonstrated that one can achieve a satisfactory description of 
the available low-energy $\Lambda N$ 
and $\Sigma N$ data within such an approach. 
First applications of the underlying hyperon-nucleon ($YN$) potential in calculations of 
binding energies for light hypernuclei were encouraging \cite{Nogga:2013,Nogga:2014}. 
In addition, and may be most remarkable, it was found that the resulting 
in-medium interaction for the $\La$ hyperon exhibits quite unusual properties.
Contrary to most phenomenological $YN$ potentials \cite{Rijken:1999,Haidenbauer:2005}, 
it becomes already repulsive at fairly low nuclear densities $\rho$, i.e.
for $\rho$ in the order of two-to-three times that of normal nuclear matter \cite{Haidenbauer:2017}. 
For such an interaction the onset for hyperon formation in neutron stars could 
be shifted to rather 
high densities, a feature that appears to be promising as a possible explanation 
for the so-called hyperon puzzle \cite{Chatterjee:2016}. 
The latter refers to the still unsolved question how one can reconcile the
softening of the equation-of-state due to the appearance of hyperons with
the observed large size (mass) of neutron stars 
\cite{Chatterjee:2016,Weissenborn:2012,Lonardoni:2015,Tolos:2017}. 

The $YN$ potential up to NLO in SU(3) chiral EFT consists of contributions from one- 
and two-pseudoscalar-meson exchange diagrams (involving the Goldstone boson octet $\pi$, $\eta$, $K$)
and  from four-baryon contact terms without and with two derivatives.
In deriving such an $YN$ potential in Ref.~\cite{Haidenbauer:2013} the SU(3) flavor 
symmetry was considered primarily as a working hypothesis 
and not so much as a fundamental prerequisite, as  emphasized in that work. 
Accordingly, the bar\-yon-baryon-meson  coupling constants for the pseudoscaler 
mesons were fixed in line with SU(3) symmetry and the symmetry 
was also exploited to derive relations between the various low-energy constants (LECs) 
that characterize the strength of the contact interactions. 
At the same time, in the actual calculation 
the SU(3) symmetry is broken, first by the mass differences between the pseudoscalar mesons
entering the potential, and second by those of the baryons ($N$, $\Lambda$, $\Sigma$) 
in the evaluation of the reaction amplitudes when solving
a coupled-channel ($\La N$-$\Si N$) scattering equation. 
For these masses the known physical values were already utilized
 in the leading-order (LO) study \cite{Polinder:2006}.

In addition, and contrary to past studies of the $YN$ interaction within phenomenological
approaches \cite{Rijken:1999,Nagels:2019}, 
no use of SU(3) symmetry was made to constrain 
the (strangeness $S=-1$) $YN$ potential by information from ($S=0$) $NN$ scattering. 
One reason for this was the observation that a combined (and realistic) description of 
the $YN$ and $NN$ systems with contact terms that fulfil strict SU(3) symmetry turned 
out to be intractable. Specifically, the friction between the strengths needed for 
reproducing the $pp$ (or $np$) $^1S_0$ phase shifts and the $\Sigma^+ p$ cross section 
could not be reconciled in a scenario which maintained SU(3) symmetry for the contact 
terms \cite{Haidenbauer:2014}. 
Another and equally important reason was the goal to explore in how far the $YN$ data 
themselves already allow one to pin down the interaction in the $S=-1$ sector. 
It should be emphasized that the aspects discussed above apply only to the interaction 
in the $S$ waves. Since there are practically no data for differential observables,
it is impossible to fix the $YN$ contact terms in the $P$-waves. In this case, 
implementing constraints from the $NN$ sector provided by SU(3) symmetry is essential, 
cf. the corresponding discussion in Ref.~\cite{Haidenbauer:2013}. 

Evidently, under the premises described above, an excellent reproduction of the 
available $YN$ data is possible, as shown in Ref.~\cite{Haidenbauer:2013}. 
Indeed, for the commonly considered set of 36 low-energy $\La N$ and 
$\Si N$ data points a $\chi^2$ of around $16$ could be achieved. This value
is comparable or even better than the results 
obtained with elaborate phenomenological models derived in 
the traditional meson-exchange  picture \cite{Rijken:1999,Nagels:2019}. 
Interestingly, it turned out that the fit to the $YN$ data allowed one to fix the majority of
the $S$-wave LECs. Nonetheless, some correlations between the values of the 
$S$-wave LECs at LO and NLO persisted, as already pointed 
out in that work. Those were attributed to the fact that the 
fitted $\Si^- p$ and $\Si^+ p$ cross sections lie all within 
a rather narrow energy interval near threshold so that 
there is only a fairly weak sensitivity to the 
momentum-dependent terms that involve the NLO LECs, 
see the appendix for explicit expressions of the contact
interaction. 
The correlations found for the $S$-wave LECs suggest that 
alternative realizations of the $\Lambda N$ and $\Sigma N$ contact interaction should be possible. 
However, in view of the excellent  $\chi^2$ obtained in the initial study \cite{Haidenbauer:2013},
at that stage, it seemed unnecessary to explore these correlations further.

In the present work, we want to catch up on this issue and consider 
variations of the $YN$ potential due to the aforementioned ambiguities
in the LECs. The questions that can be addressed in this way are: 
(i) Is it possible to achieve a description of $\Lambda N$ and $\Sigma N$
scattering for an alternative set of LECs that is comparable or even better 
than the one in Ref.~\cite{Haidenbauer:2013},
i.e. with comparable or even lower $\chi^2$? 
(ii) Do the resulting $\La N$ and $\Si N$ potentials have different properties?
In particular, do they lead to qualitatively different results when employed in 
studies of few- and many-body systems involving hyperons?

One possibility to eliminate the aforementioned correlations between the 
LECs consists in implementing additional constraints to simply reduce
the number of contact terms that need to be fitted to the $YN$ data. 
A sensible choice is to impose SU(3) symmetry more strictly than 
in Ref.~\cite{Haidenbauer:2013} and to take into account the symmetry 
relations between $YN$ and $NN$ also for the $S$-waves, and not only 
for the $P$-waves. 
How this can be done in practice was demonstrated in Ref.~\cite{Haidenbauer:2014}
for a specific case, namely the $^1S_0$ partial wave in the $NN$, $\Si N$, and
$\Si\Si$ systems. This work exploited the fact that at NLO in the perturbative 
expansion of the baryon-baryon potentials genuine SU(3) symmetry-breaking 
contact terms arise \cite{Petschauer:2013}.
Accordingly, the LO LECs for $NN$ and $YN$ $S$-waves are no longer 
completely constrained by SU(3) symmetry, only those at NLO. 
This allows one to remedy the friction between the $pp$ and $\Si^+ p$
results mentioned above and, at the same time, stay in line with the
underlying power counting of SU(3) chiral EFT. 
In the present work, we now apply this scheme to all $S$-waves 
of the $NN$, $\La N$, and $\Si N$ systems. 

Anticipating our results, it turns out that an equally convincing
description of $\Lambda N$ and $\Sigma N$ scattering data can be
achieved based on such an alternative choice of the LECs. Indeed,
the cross sections (actually all considered two-body observables)
are practically indistinguishable from those in 
Ref.~\cite{Haidenbauer:2013}.
Small variations are observed for the predicted binding energies
for the hypertriton $^3_\La \rm H$ and the 
$^4_\La \rm H$ and $^4_\La \rm He$ hypernuclei. 
However, in case of the properties of the hyperons in nuclear
matter, the differences are much more sizable. Specifically,
the in-medium interaction of the $\Lambda$ predicted by the
new potential is now considerably more attractive and becomes
repulsive at much higher nuclear densities as compared to the EFT
interaction published in~\cite{Haidenbauer:2013}. 

The paper is structured in the following way: 
In the next section, a summary of the formalism is provided. Since
a thorough description of the approach for treating $YN$ scattering within
SU(3) chiral EFT is available in Ref.~\cite{Haidenbauer:2013}, 
we will be brief here. Details that are needed to understand in how
far the EFT interaction proposed in the present work differs from that in \cite{Haidenbauer:2013} are summarized in an appendix. 
The coverage of the Brueckner reaction-matrix formalism that is employed 
for evaluating the in-medium properties of the $\La$ and $\Si$ is likewise
kept short. Here, we refer the reader to Refs.~\cite{Reuber:1994} and
\cite{Haidenbauer:2015} for details. 
In Sect.~\ref{sec:Results}, the results for the alternative potential are
presented  and compared to the ones published in 2013 
(for $\La N$, $\Si N$ scattering) and 2015 (for nuclear matter).  
Implications of our results are discussed in Sect.~\ref{sec:Discussion}. 
The paper ends with concluding remarks. 


\section{Formalism} 
\label{sec:Formalism}

\subsection{{\boldmath$\La N$} and {\boldmath$\Si N$} scattering}
The derivation of the chiral baryon-baryon potentials for the strangeness sector
using the Weinberg power counting is outlined in Refs.
\cite{Haidenbauer:2013,Polinder:2006,Petschauer:2013}.
The LO potential consists of four-baryon contact terms without derivatives and of
one-pseudoscalar-meson exchanges while at NLO contact terms with two derivatives
arise, together with contributions from (irreducible) two-pseudoscalar-meson exchanges.
The contributions from pseudoscalar-meson exchanges (the Goldstone bosons $\pi$, $\eta$, $K$
of the spontaneously broken chiral symmetry of QCD) are completely
fixed by the assumed SU(3) flavor symmetry.
On the other hand, the strength parameters associated with the contact terms, the low-energy
constants (LECs), need to be determined in a fit to data. How this is done is described in
detail in Ref.~\cite{Haidenbauer:2013}. With regard to the alternative version considered 
in the present work, the strategy followed is described in the beginning of 
Sect.~\ref{sec:Results}. 
Note that, in general, SU(3) symmetry is also imposed for the contact terms which
reduces the number of independent LECs that can contribute.

After a partial-wave projection \cite{Polinder:2006}, the potential $V$ is
inserted into a regularized coupled-channels Lippmann-Schwinger (LS) equation
for the $YN$ $T$-matrix $T$, 
\begin{eqnarray}
&&T^{\kappa''\kappa',J}_{\nu''\nu'}(p'',p';\sqrt{s})=
V^{\kappa''\kappa',J}_{\nu''\nu'}(p'',p')+
\nonumber\\&&
\sum_{\kappa,\nu}\int_0^\infty \frac{dp \, p^2}{(2\pi)^3} \, V^{\kappa''\kappa,J}_{\nu''\nu}(p'',p)
\frac{2\mu_{\nu}}{q_{\nu}^2-p^2+i\eta}T^{\kappa\kappa',J}_{\nu\nu'}(p,p';\sqrt{s})\ . \nonumber \\
&&
\label{LS} 
\end{eqnarray}
and its solution provides us the reaction amplitudes.
The label $\nu$, $\nu'$, and $\nu''$ in Eq.~(\ref{LS}) indicate the particle channels 
and the label $\kappa$, $\kappa'$, and $\kappa''$ the partial wave ones where $J$ is the total angular momentum \cite{Polinder:2006}. 
$\mu_\nu$ is the pertinent reduced mass. The on-shell momentum in the 
intermediate state, $q_{\nu}$, is defined by 
$\sqrt{s}=(m^2_{B_{1,\nu}}+q_{\nu}^2)^{1/2}+(m^2_{B_{2,\nu}}+q_{\nu}^2)^{1/2}$.
Relativistic kinematics is used for relating the laboratory energy $T_{{\rm lab}}$ of the hyperons
to the c.m. momentum. Otherwise, we use non-relativistic kinematics for the solution of the
two- and more-baryon equations. 

We solve the LS equation in the particle basis in order to incorporate the correct physical
thresholds. The Coulomb interaction is taken into account appropriately via the Vincent-Phatak
method \cite{VP}. 
Regularization is done in the same way as in our initial work \cite{Haidenbauer:2013}, 
see also Ref.~\cite{Epelbaum:2005}. This means that the potentials in the LS
equation are cut off with an exponential regulator function, $f_R(\Lambda) =
\exp\left[-\left(p'^4+p^4\right)/\Lambda^4\right]$, so that high-momentum components 
are removed \cite{Epelbaum:2005}. We consider cutoff values in the range
$\Lambda=500$ -- $650$ MeV where the best $\chi^2$ values were achieved in
the 2013 study \cite{Haidenbauer:2013}. 
As before, we present our results as bands which reflect the variation with the 
cutoff and, thus, indicate a lower bound for the uncertainty due to truncation of the chiral expansion.
A more sensible way for estimating this uncertainty, that does not
rely on  cutoff variation, has been proposed in
Refs.~\cite{Epelbaum:2015,Binder:2015mbz}
and we will show selected results based on that method, too. 
However, one should keep in mind that the present $YN$ 
interactions are still only on the level of NLO which possibily leads to an 
underestimation of the uncertainty (as explained in more detail below). 

\subsection{{\boldmath$\La $} and {\boldmath$\Si $} in nuclear matter}
\label{sec:MatterF} 
The nuclear matter properties of the $\La$ and $\Si$ hyperons are evaluated within the conventional
Brueckner theory. We summarize below only the essential elements. A detailed description
of the formalism can be found in
Refs.~\cite{Reuber:1994,Haidenbauer:2015}, see also Ref.~\cite{Vid00}.
We consider a $\La$ or $\Si$ hyperon with momentum ${\vec p}_Y$ in nuclear matter at density $\rho$.
In order to determine the in-medium properties of the hyperon, we employ the Brueckner
reaction-matrix formalism and calculate  the $Y N$ reaction matrix $G_{Y N}$, defined by the
Bethe-Goldstone equation
\begin{eqnarray}
\nonumber
&&\langle Y N | G(\zeta) | Y N \rangle = \langle Y N | V | Y N \rangle 
\\
\nonumber
&&+ \sum_{Y'N} \ \langle Y N | V | {Y'N} \rangle \,
\langle {Y'N} | \frac{Q}{\zeta - H_0}|{Y'N} \rangle \, \langle {Y'N} | G(\zeta) | Y N \rangle , 
\\
&&
\label{Eq:G1}
\end{eqnarray}
with $Y,Y'$ = $\La$, $\Si$.
Here, $Q$ denotes the Pauli projection operator  which excludes intermediate
$Y N$-states with the nucleon inside the Fermi sea. $H_0$ is the kinetic energy of the $YN$ system. 
The starting energy $\zeta$ for an initial $Y N$-state with momenta ${\vec p}_Y $ and
${\vec p}_N$ is given by
\begin{equation}
\zeta = E_Y (p_Y) + E_N (p_N),
\end{equation}
where the single-particle energy $E_\alpha (p_\alpha)$ ($\alpha = \La, \Si, N$)
includes not only the (non-relativistic) kinetic energy and the baryon mass but
in addition the single-particle (s.p.) potential $U_\alpha (p_\alpha, \rho)$:
\begin{equation}
E_\alpha (p_\alpha) = m_\alpha + \frac{\vec p^{\,2}_\alpha}{2m_\alpha} + U_\alpha (p_\alpha,
\rho)\, .
\label{Eq:G2}
\end{equation}
The so-called gap-choice \cite{Reuber:1994} for the intermediate-state spectrum is adopted.
The $Y$ single-particle potential  \\ {$U_Y(p_Y,\rho)$
is given by the following integral and sum over diagonal $Y N$ $G$-matrix elements:
\begin{equation}
U_Y(p_Y,
\rho) = \int\limits_{|\vec p_N|< k_F}  \frac{d^3p_N}{(2\pi)^3}\, \rm{Tr}
\langle {\vec p}_Y ,{\vec p}_N | G_{Y N}(\zeta) | {\vec p}_Y, {\vec p}_N \rangle 
\,,
\label{Eq:G3}
\end{equation}
where  $\rm{Tr}$ denotes the trace in spin- and isospin-space.
Note that $\rho = 2k^3_F/3\pi^2$ for symmetric nuclear matter and
$\rho = k^3_F/3\pi^2$ for neutron matter, where $k_F$ denotes the Fermi momentum. 

Eqs.\,(\ref{Eq:G1}) and (\ref{Eq:G3}) are solved self-consistently in a standard way,
with  $U_Y(p_Y, \rho)$ appearing also in the starting energy $\zeta$. Like in
Ref.~\cite{Haidenbauer:2017}, the nucleon single-particle potential
$U_N(p_N,\rho)$ is taken from a calculation of nuclear matter employing a
phenomenological $NN$ potential. Specifically, we resort to results for the Argonne $v_{18}$
potential published in Ref.\,\cite{Isaule:2016}.
As pointed out in Ref.~\cite{Reuber:1994}, calculations of hyperon potentials in nuclear
matter using the gap-choice are not too sensitive to the details of $U_N(p_N,\rho)$. 
Indeed, the difference for, e.g., $U_\La(0,\rho)$ using $U_N(p_N,\rho)$ from 
Ref.~\cite{Isaule:2016} or the parameterization utilized in
Ref.~\cite{Haidenbauer:2015} amounts to around $1$~MeV at nuclear matter saturation
density $\rho= 0.17$~fm$^{-3}$ ($k_F=1.35$~fm$^{-1}$).

Since, at this stage, we are primarily interested in
comparing the results for the two $YN$ interactions, 
we refrain from a much more time-consuming calculation necessitated by the  so-called continuous choice
\cite{Petschauer:2016}.

\subsection{Faddeev and Yakubovsky equations}

Binding energies of light $A=3$ and $4$ hypernuclei can 
be obtained by solving Faddeev or Yakubovsky equations in 
momentum space 
\cite{Miyagawa:1993,Miyagawa:1995,Noggaphd:2001,Nogga:2002}. 
The method is well suited for chiral $YN$ interactions since it allows one to employ
non-local interactions including particle and partial wave channel couplings. Indeed, 
the works by Miyagawa et al.~\cite{Miyagawa:1993,Miyagawa:1995} constitute the first
successful attempt to use realistic meson-exchange potentials (including tensor forces and the 
$\Lambda N$-$\Sigma N$ coupling) directly in a hypertriton calculation within the Faddeev framework. 
In Ref.~\cite{Miyagawa:1993}, an overview of earlier
calculations of the hypertriton is provided. 

The few-body results given in this work have been obtained 
solving the three- and four-body non-relativistic Schr\"odinger equation 
in momentum space by rewriting them into Faddeev or Yakubovsky equations, 
respectively. For a bound state of one hyperon and two nucleons, one finds 
two coupled Faddeev equations 
\begin{eqnarray}
\psi_1 & = & G_0 T_{NN} (1-P_{12}) \psi_2 \nonumber \\[4pt] 
\psi_2 & = & G_0 T_{YN} (\psi_1-P_{12}\psi_2) \ , 
\label{3BE} 
\end{eqnarray}
for the two independent Faddeev components $\psi_{1}$ and $\psi_{2}$ \cite{Miyagawa:1993,Noggaphd:2001}.
$P_{12}$ is the operator permuting the coordinates of nucleon 1 and 2. The $NN$ and $YN$ interactions
enter via the corresponding $T$-matrices $T_{NN}$ and $T_{YN}$.
They are obtained by solving LS equations embedded in the three- or four-body 
Hilbert space and which are therefore fully off-shell. 
The free propagator is denoted by $G_0$. We are only considering bound states. 
Therefore, directly using the Schr\"odinger equation is in principle possible. However,
using two different kinds of Jacobi coordinates for $\psi_1$ and $\psi_2$ 
that single out either an $NN$ or a $YN$ subsystem leads to an improved convergence with
respect to partial waves. Therefore, the rewriting in Faddeev 
equations is advisable also for a bound state calculation. 
The two basis sets used for the calculation are then denoted by 
\begin{eqnarray}
& & | p_{12} p_3 \alpha_1 \rangle \cr 
& & = | p_{12} p_3 ((l_{12} s_{12}) j_{12} (l_3 \frac{1}{2}) I_3) J; (t_{12} t_Y)T \rangle \cr 
& & | p_{23} p_1 \alpha_2 \rangle \cr 
& & = | p_{23} p_1 ((l_{23} s_{23}) j_{23} (l_1 \frac{1}{2}) I_1) J; (t_{23} \frac{1}{2})T \rangle \ , \cr 
&&
\end{eqnarray}
where $p_{ij}$ are the magnitudes of the pair momenta and  $p_k$ is the magnitude 
of the spectator momentum. Their angular dependence is expanded in orbital angular momenta  $l_{ij}$ and 
$l_k$, respectively. The orbital angular momenta are coupled with the spin of the 
pair $s_{ij}$ and the spin of the spectator baryon to the intermediate 
angular momentum $j_{ij}$ and $I_k$. These are then coupled to the total 
angular momentum of the hypertriton $J=\frac{1}{2}$. Since we work in the isospin 
basis, the pair isospin $t_{ij}$ is either coupled with the isospin $t_Y$ of the 
spectator hyperon or the isospin $\frac{1}{2}$ of the spectator nucleon 
to total isospin $T=0$. The number of partial waves is finite, if one restricts 
$j_{ij} \le j_{\rm max}$. For the calculations shown here, we chose $j_{\rm max}=6$. The 
binding energies are then converged to better than 1~keV. The momenta 
are discretized typically using a grid of 44 or 58 points for $p_{ij}$ and $p_k$,
respectively. 

In the four-body case, we find a set of five Yakubovsky equations 
for five independent Yakubovsky components $\psi_{1A}$ $\psi_{1B}$, $\psi_{1C}$, 
$\psi_{2A}$, and $\psi_{2B}$
\begin{eqnarray}
\psi_{1A} & = & G_0 T_{NN} P (\psi_{1A}+\psi_{1B}+\psi_{2A}) \nonumber \\[4pt]
\psi_{1B} & = & G_0 T_{NN} ((1-P_{12})(1-P_{23})\psi_{1C}+ P \psi_{2B})  \nonumber \\[4pt]
\psi_{1C} & = & G_0 T_{YN} (\psi_{1A}+\psi_{1B}+\psi_{2A} \cr & &
                            -P_{12}\psi_{1C}+P_{12}P_{23}\psi_{1C}  + P_{13}P_{23}\psi_{2B})  \nonumber \\[4pt]
\psi_{2A} & = & G_0 T_{NN} ((P_{12}-1)P_{13})\psi_{1C}+\psi_{2B}) \nonumber \\[4pt]
\psi_{2B} & = & G_0 T_{YN} P (\psi_{1A}+\psi_{1B}+\psi_{2A}) \ . 
\label{4BE} 
\end{eqnarray}
Again, all five components are expanded in a different set of Jacobi basis states. 
The first three components use so-called $3+1$ coordinates, which single out 
one pair momentum $p_{ij}$, one baryon moving relative to the pair with momentum $p_{k}$ and the spectator moving 
relative to the other three baryons with momentum $q_l$. The other two components use 
$2+2$ coordinates which single out two pair momenta $p_{ij}$ and $p_{kl}$ and 
a momentum $q$ describing the relative motion of the two pairs. The angular dependence 
is expanded in terms of corresponding orbital angular momenta. These are coupled with 
spins to a total angular momentum $J$. Similarly, the isospins are finally coupled to 
total isospin $T$. More details are given in~\cite{Noggaphd:2001}. In the case 
of the four-body system, restricting pair angular momenta $j_{ij} \le j_{\rm max}$ 
is not sufficient to get a finite set of equations. We therefore 
impose two more constraints: all orbital angular momenta $l \le 4$ and the sum 
of all three orbital angular momentum quantum numbers is  less than $l_{\rm sum}=8$. 
We carefully checked that the energies are converged to better than 10~keV 
for the chiral $NN$ and $YN$ interactions. For the three different 
momenta, a discretization using 52, 56 and 48 grid points is typically sufficient 
to get an accuracy better than a few keV.  Since phenomenological interactions 
are less soft, we use more partial waves in this case in order to get a similar accuracy.
We note that the number of partial wave and isospin channels are especially larger 
for the excited $J^\pi=1^+$ state. With the restrictions defined above, we had to 
take more than 13000 partial wave combinations into account. The discretized set of 
linear equations is than more than $10^9$ dimensional. However, the accuracy of 10~keV 
that we obtain for the four-body system is sufficient to discuss the $NN$ and $YN$ 
interaction dependence of our results. 

\section{Results} 
\label{sec:Results}

In the following subsections, we present results for $\La N$ and $\Si N$
scattering, for binding energies of light hypernuclei, and for the
$\La$ and $\Si$ s.p. potentials in nuclear matter for our NLO 
chiral EFT interactions. Thereby, we will show results for two different 
fits. We refer to the original NLO fit \cite{Haidenbauer:2013} by NLO13. 
Additionally, we devised a new version in the way described below which 
we will label NLO19 in the following. 
Occasionally, also results for $YN$ potentials based on the traditional 
meson-exchange approach \cite{Rijken:1999,Haidenbauer:2005} 
will be shown for illustration. 

Let us first describe in more detail how the contact terms for the
new $YN$ interaction NLO19 were fixed.
The total number of independent LECs up to NLO amounts to $13$ for the $^1S_0$ and 
$^3S_1$-$^3D_1$ $YN$ partial waves, with $4$ ($6$) for the singlet (triplet) $S$-states 
and $3$ for the $^3S_1$$\leftrightarrow$$^3D_1$ transition, see the appendix. 
In Ref.~\cite{Haidenbauer:2013}, their values have been fixed 
by a fit to the usually considered \cite{Haidenbauer:2013,Rijken:1999,Haidenbauer:2005,Polinder:2006}  
set of low-energy $\La p$, $\Si^- p$, and $\Si^+ p$ data
\cite{Sec68,Ale68,Eng66,Eis71,Hep68}. 
Now, with SU(3) symmetry imposed, three of those can be inferred from the $NN$ interaction,
namely $C^{27}_{^1S_0}$, $C^{10^*}_{^3S_1}$, and $C^{10^*}_{^3S_1-^3D_1}$, so that
there are only $10$ LECs left to be fitted. 
For the refit NLO19 of the present work, we utilize the $NN$ values of Ref.~\cite{Petschauer:2016}. 
There, an $NN$ potential has been established within the same framework and the same 
regularization scheme and, therefore, allows one to enforce SU(3) symmetry 
of the LECs. The pertinent LECs were determined by a fit to $np$ phase shifts. 
$pp$ phase shifts could be used instead for $^1S_0$ partial wave. 
The difference in the corresponding LEC $C^{27}$ is, however, tiny, see
Ref.~\cite{Haidenbauer:2014}, so that we ignore this for the time being. 

We use the NLO potential from 2013 \cite{Haidenbauer:2013} as starting point for
our new fit, of course, with substituting the LECs that are fixed from the $NN$ 
sector. Remarkably, in case of the $^1S_0$, there is only a small difference 
in the actual values for the $C^{27}$'s found in the fit to the $YN$ data 
(cf. Table~3 in Ref.~\cite{Haidenbauer:2013}) and the ones from the $NN$ results 
(cf. the appendix). These coincide within $5-10$~\%, despite of being fitted 
independently. One could interpret this as a sign that, beyond corrections at 
leading order, SU(3) symmetry is fairly well realized.

In case of the $^3S_1$, the situation is different. Here the values for $C^{10^*}$
from the fit to the $np$ phase shifts are more than a factor $5$ smaller than those determined
in the $YN$ study. Indeed,  the LECs for the new fits are now well in 
line with being of ``natural size'' \cite{Epelbaum:2006}. 
Note that the LECs for the
${10}$ and ${10^*}$ representations have been erroneously interchanged in
Table~3 of Ref.~\cite{Haidenbauer:2013}!  We view these large variations primarily as a sign of
the correlations between the LO and NLO LECs discussed already in the introduction. 
 
The best description of $YN$ data was obtained for the range $500 - 650$~MeV for 
the cutoff $\Lambda$ in the regulator function in Ref.~\cite{Haidenbauer:2013}. 
Therefore, we consider again this range in the present work. 
For smaller values, there is a 
rapid deterioration in the $\chi^2$, cf. Table~5 in \cite{Haidenbauer:2013},
and likewise (though less dramatic) for larger values.    

All LECs in the $P$-waves are taken over from Ref.~\cite{Haidenbauer:2013}. No
readjustment is done in this case. Thus, all differences in the results for
the two interactions reported below stem from the differences in the $S$-waves. 

\subsection{{\boldmath$\La N$} and {\boldmath$\Si N$} scattering}
\label{sec:Scattering} 

\begin{table*}[th]
\caption{Scattering lengths ($a$) and effective ranges ($r$) for singlet (s) and 
  triplet (t) $S$ waves, for $\Lambda p$, $\Si N$ with isospin $I=1/2$,
  and $\Si^+ p $ ($I=3/2$).  
  In addition, the achieved $\chi^2$ for the 36 data points is listed. 
  $^*$In case of the J\"ulich '04 potential, the capture ratio was not included in 
  the fit and the evaluation of the $\chi^2$. 
  }
 \label{tab:R0}
\vskip 0.1cm
\renewcommand{\arraystretch}{1.4}
\begin{center}
\begin{tabular}{|c||rrrr|rrrr||rr|}
\hline
\hline
& \multicolumn{4}{|c|}{NLO13} & \multicolumn{4}{|c||}{NLO19} & \ J\"ulich '04 \ & \ NSC97f \ \\
\hline
${\Lambda}$ [MeV] & 500 & 550 & 600 & 650 & 500 & 550 & 600 & 650  & & \\
\hline
$a^{\La p}_s$ & $-2.91$ & $-2.91$ & $-2.91$ & $-2.90$ & $-2.91$ & $-2.90$ & $-2.91$ & $-2.90$ 
& $-2.56$ & $-2.60$ \\
$r^{\La p}_s$ & $ 2.86$ & $ 2.84$ & $ 2.78$ & $ 2.65$ & $ 3.10$ & $ 2.93$ & $ 2.78$ & $ 2.65$ 
& $ 2.74$ & $ 3.05$ \\
\hline
$a^{\La p}_t$ & $-1.61$ & $-1.52$ & $-1.54$ & $-1.51$ & $-1.52$ & $-1.46$ & $-1.41$ & $-1.40$ 
& $-1.67$ & $-1.72$ \\
$r^{\La p}_t$ & $ 3.05$ & $ 2.83$ & $ 2.72$ & $ 2.64$ & $ 2.62$ & $ 2.61$ & $ 2.53$ & $ 2.59$ 
& $ 2.93$ & $ 3.32$ \\
\hline
\hline
Re\,$a^{\Si N}_{s}$ &  $ 1.00$ & $ 0.98$ & $ 0.90$ & $ 0.87$ & $ 0.99$ & $ 0.98$ & $ 0.90$ & $ 0.87$ 
& $ 0.90$ & $ 1.16$ \\
Im\,$a^{\Si N}_{s}$ &  $ 0.00$ & $ 0.00$ & $ 0.00$ & $ 0.00$ & $ 0.00$ & $ 0.00$ & $ 0.00$ & $ 0.00$ 
& $-0.13$ & $ 0.00$ \\
\hline
Re\,$a^{\Si N}_{t}$ & $ 2.61$ & $ 2.44$ & $ 2.27$ & $ 2.06$ & $ 0.95$ & $ 0.98$ & $ 2.29$ & $ 1.95$ 
& $-3.83$ & $ 1.68$ \\
Im\,$a^{\Si N}_{t}$ & $-2.89$ & $-3.11$ & $-3.29$ & $-3.59$ & $-4.77$ & $-4.59$ & $-3.39$ & $-3.85$ 
& $-3.01$ & $-2.35$ \\
\hline
\hline
$a^{\Si^+ p}_s$ &  $-3.59$ & $-3.60$ & $-3.56$ & $-3.46$ & $-3.90$ & $-3.79$ & $-3.62$ & $-3.43$ 
& $-3.60$ & $-4.35$ \\
$r^{\Si^+ p}_s$ &  $ 3.59$ & $ 3.56$ & $ 3.54$ & $ 3.53$ & $ 3.55$ & $ 3.50$ & $ 3.50$ & $ 3.52$ 
& $ 3.24$ & $ 3.16$ \\
\hline
$a^{\Si^+ p}_t$ & $ 0.49$ & $ 0.49$ & $ 0.49$ & $ 0.48$ & $ 0.42$ & $ 0.43$ & $ 0.47$ & $ 0.48$ 
& $ 0.31$ & $-0.25$ \\
$r^{\Si^+ p}_t$ & $-5.18$ & $-5.03$ & $-5.08$ & $-5.41$ & $-6.45$ & $-6.49$ & $-5.77$ & $-5.69$ 
& $-12.2$ & $-28.9$ \\
\hline
\hline
$\chi^2$     & $16.8 $ & $15.7 $ & $16.2 $ & $16.6 $ & $18.1 $ & $17.4 $ & $16.0 $ & $16.1 $ 
& $22.1^*$ & $16.7$ \\
\hline
\hline
\end{tabular}
\end{center}
\renewcommand{\arraystretch}{1.0}
\end{table*}

In this subsection, we present results for $\La N$ and $\Si N$ scattering.
In particular, we compare the results obtained with the new procedure to
those from 2013.  
A summary of the $\Lambda p$ effective range parameters
is given in Table~\ref{tab:R0} together with information about the
achieved overall $\chi^2$. The latter, listed at the bottom, provides
clear evidence that the quality of description of the $Y N$ data by the 
two interactions is identical. The differences in the $\chi^2$ 
are marginal considering the inherent residual regulator
dependence in both cases. We observe though that the dependence of the $\chi^2$ on the cutoff
is slightly different for the two interactions. The effective range parameters in the singlet state are
practically identical. Noticeable variations occur only in 
the effective range at the lower end of the considered cutoff
range. In the triplet $S$-wave, the scattering lengths differ 
in average by $7$~\%. 

\begin{figure*}[t]
\begin{center}
\includegraphics[height=100mm]{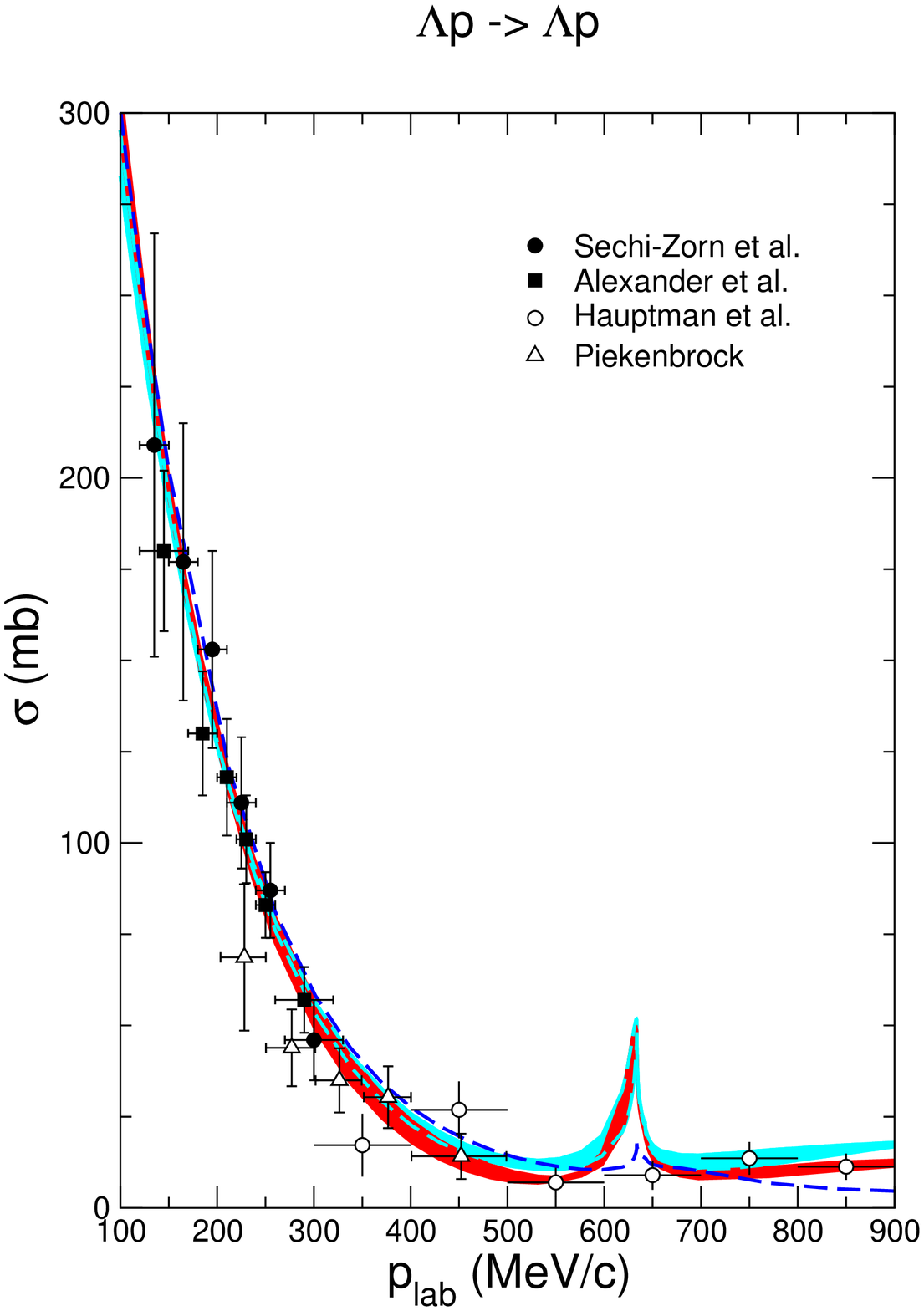}
\includegraphics[height=100mm]{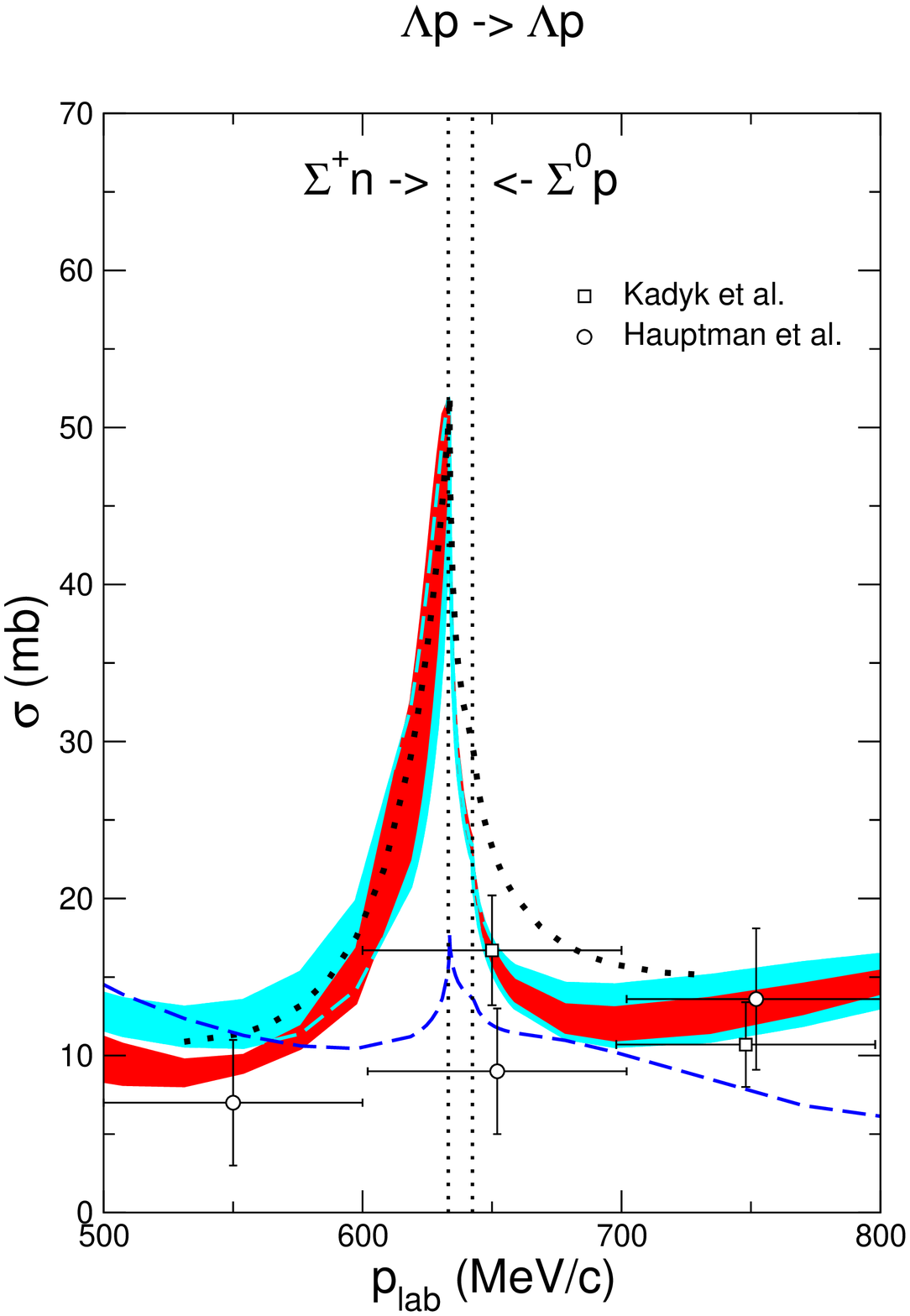}
\caption{Cross section for $\La p$ scattering as a function of $p_{lab}$.  
The red (dark) band represents the result for NLO13 
\cite{Haidenbauer:2013} including 
cutoff variations, the cyan (light) band that for the 
alternative version NLO19. 
The dashed curve is the result of the J{\"u}lich '04 meson-exchange model 
\cite{Haidenbauer:2005}, the dotted curve that of the Nijmegen NSC97f potential 
\cite{Rijken:1999}. 
The experimental cross sections are taken from Refs.~\cite{Sec68} (filled circles), \cite{Ale68} (filled squares), 
\cite{Piekenbrock,Herndon:19672} (open triangles),  
\cite{Kad71} (open squares), and \cite{Hau77} (open circles). 
The dotted vertical lines labeled with $\Si^{+}$n and $\Si^0$p indicate the thresholds of the pertinent $\Si$N channels. 
}
\label{fig:C1}
\end{center}
\end{figure*}

Results for the $\Lambda p$ cross section are displayed in Fig.~\ref{fig:C1},
where the region around the $\Si N$ thresholds is shown separately so that one
can see the details. 
As usual, the results are presented as bands that reflect the variation with
the cutoff $\Lambda$. The results for NLO13 are shown as red (dark) bands 
while the new results are shown as cyan (light) bands. In this figure and the following ones,  
$YN$ data included in the fitting procedure
\cite{Sec68,Ale68,Eng66,Eis71}
are displayed by filled symbols, while for additional data at
higher energies 
\cite{Piekenbrock,Herndon:19672,Kad71,Hau77,Ste70,Kon00,Ahn05}
open symbols are used. 

\begin{figure*}
\begin{center}
\includegraphics[height=100mm]{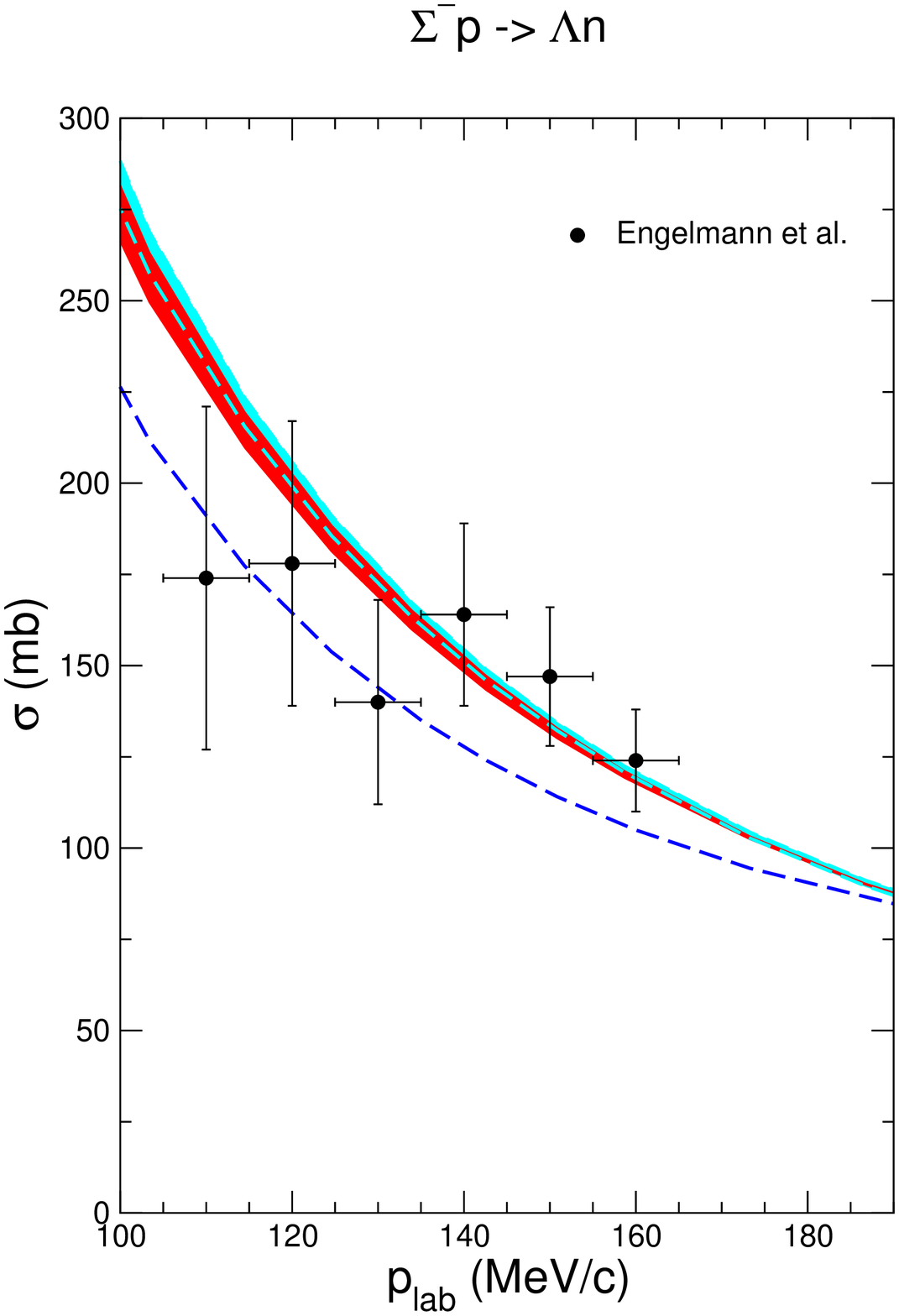}
\includegraphics[height=100mm]{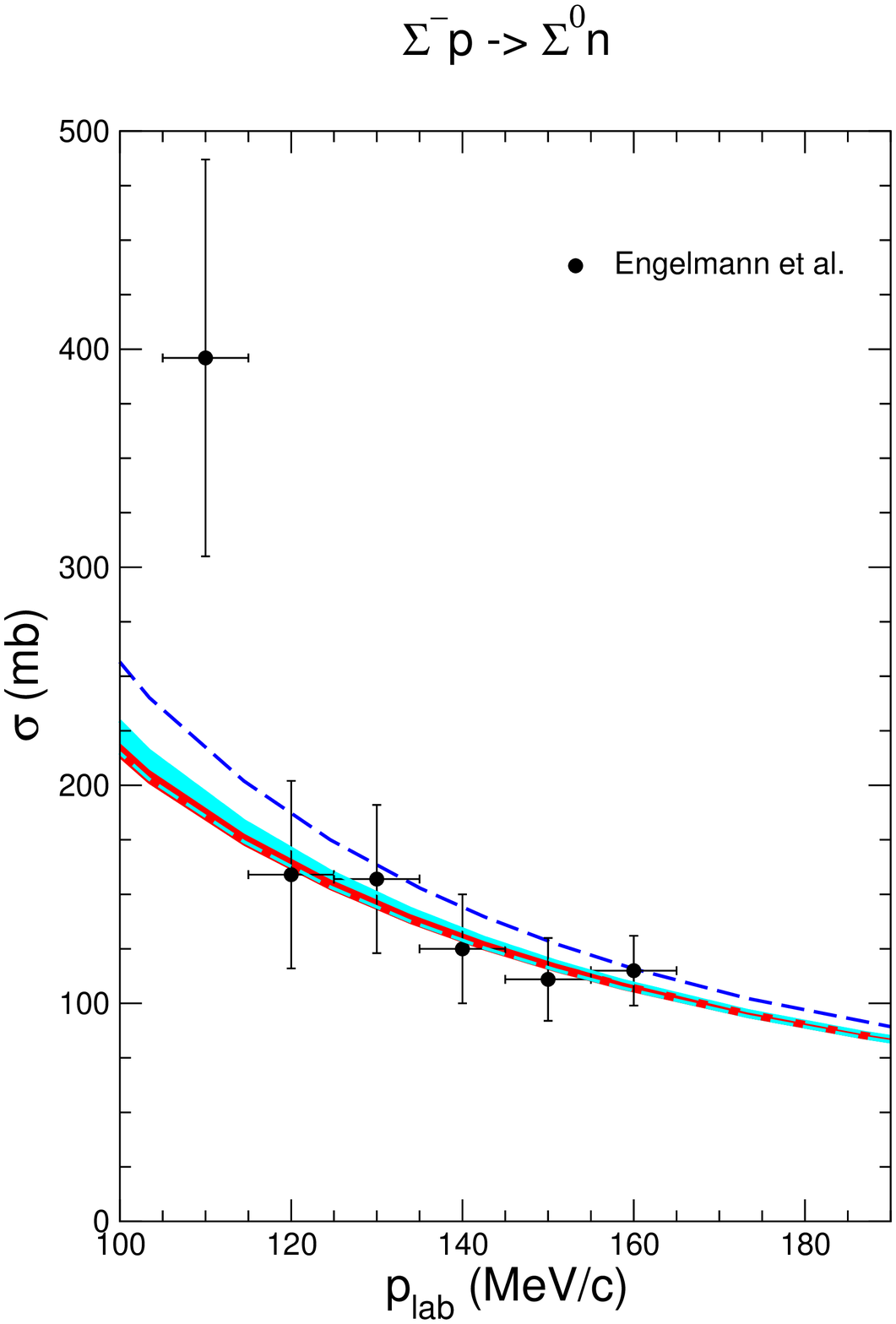}

\includegraphics[height=100mm]{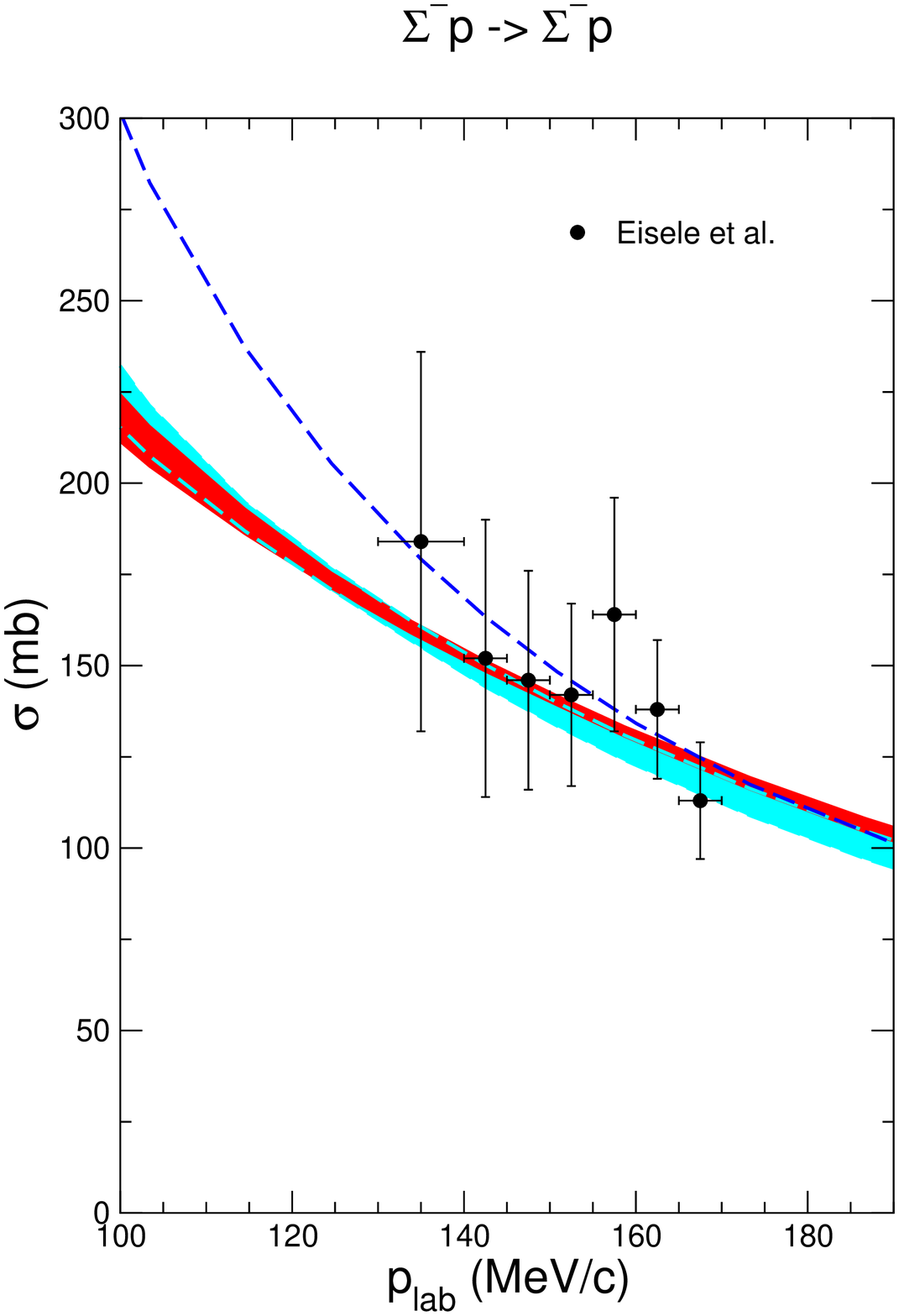}
\includegraphics[height=100mm]{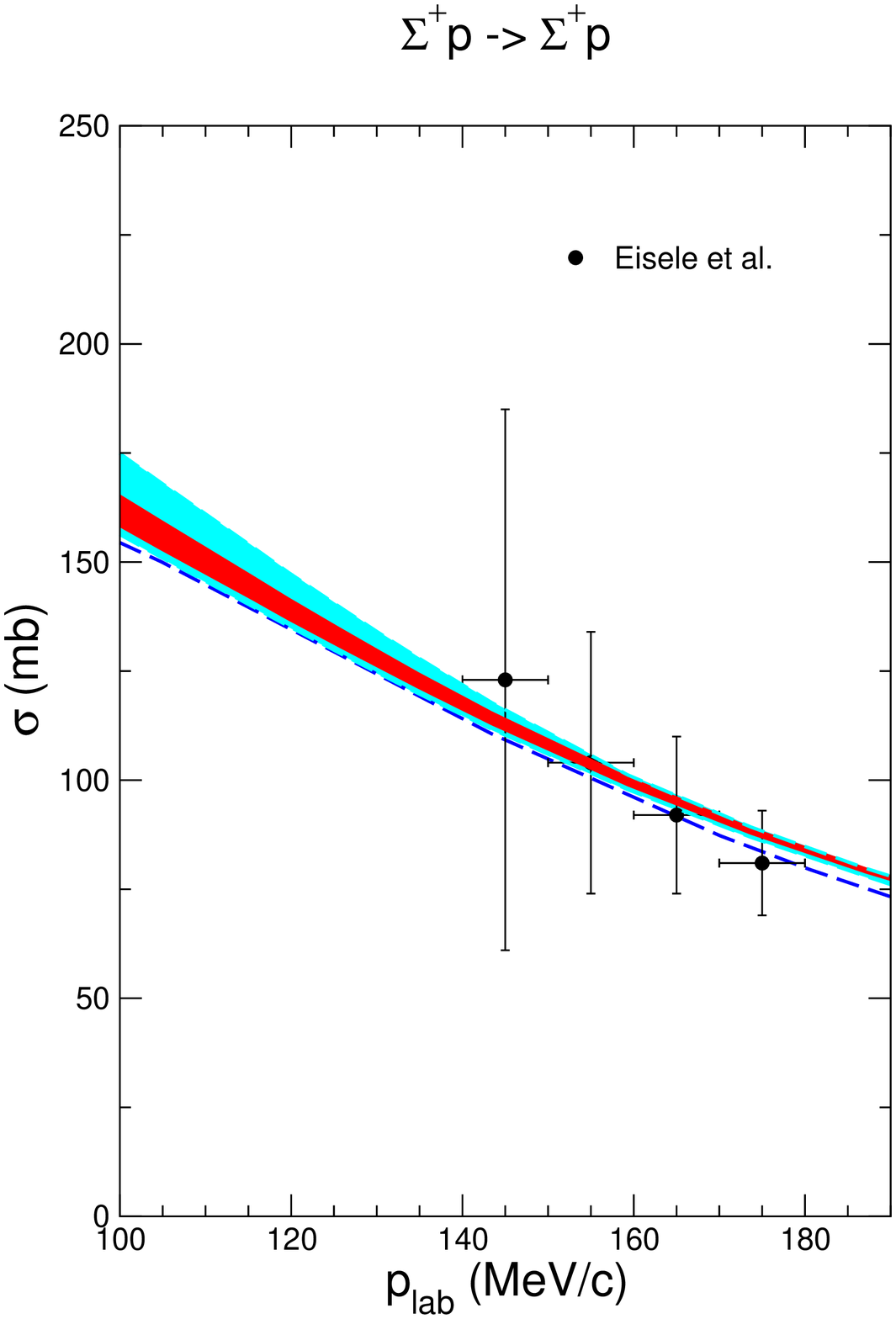}
\caption{Different $\Si$N and $\Si$N $\to$ $\La$N cross sections.
Same description of curves as in Fig.~\ref{fig:C1}.
The experimental cross sections are taken from Refs.~\cite{Eng66} 
($\Sigma^-p \to \Lambda n$, $\Sigma^-p \to \Sigma^0 n$) 
and \cite{Eis71} 
($\Sigma^-p \to \Sigma^- p$, $\Sigma^+p \to \Sigma^+ p$).
}
\label{fig:C2}
\end{center}
\end{figure*}

\begin{figure*}
\begin{center}
\includegraphics[height=100mm]{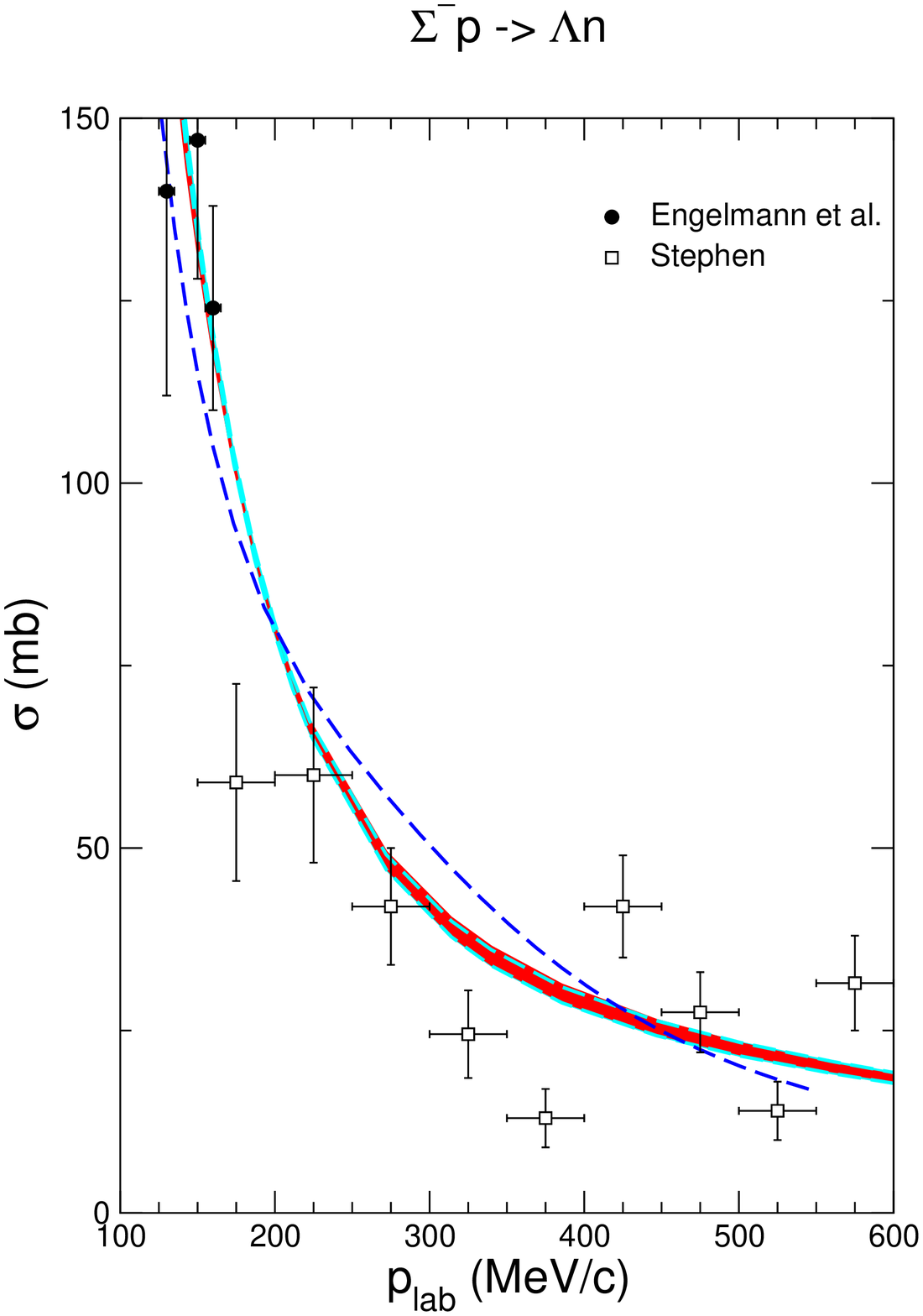}
\includegraphics[height=100mm]{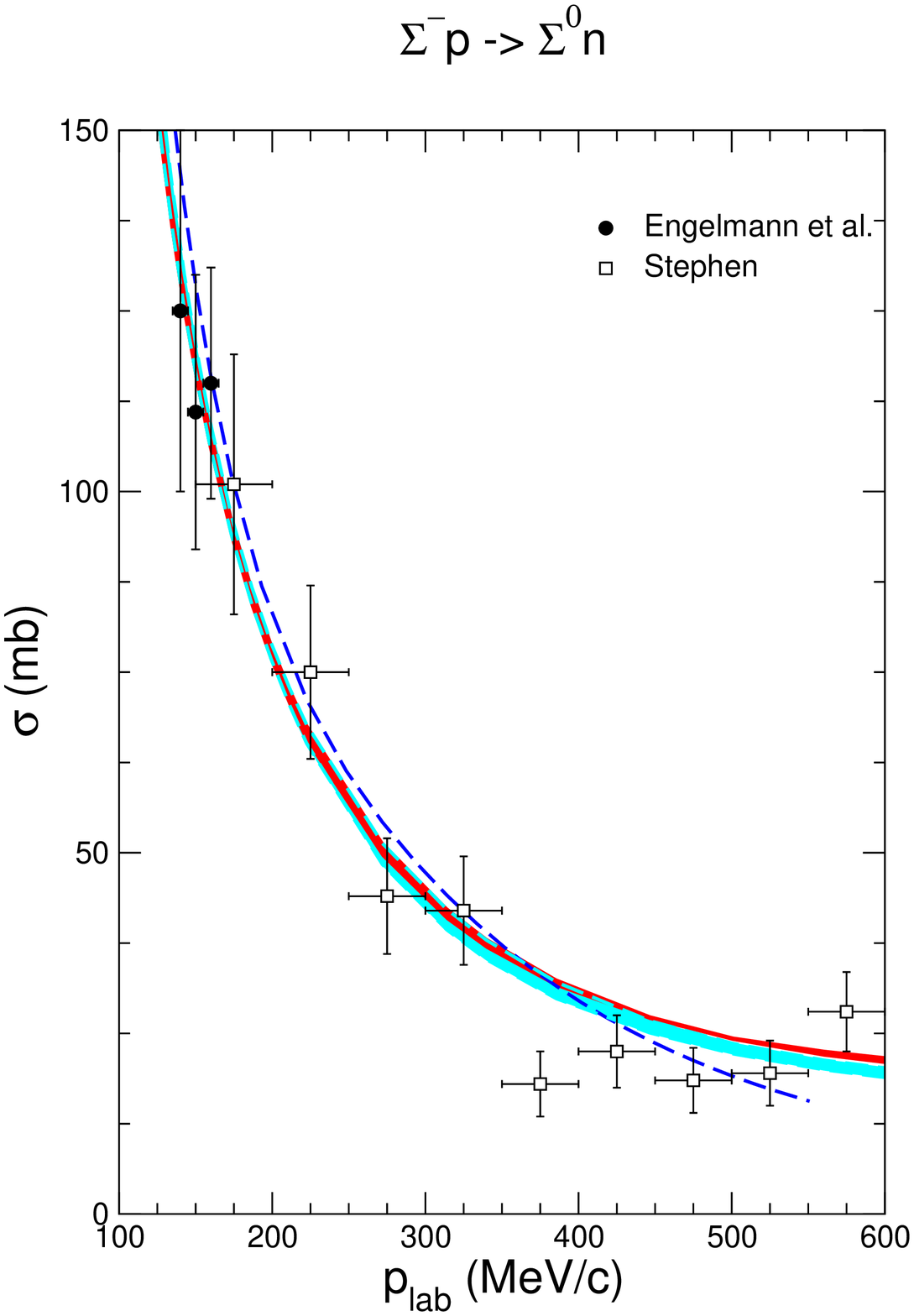}

\includegraphics[height=100mm]{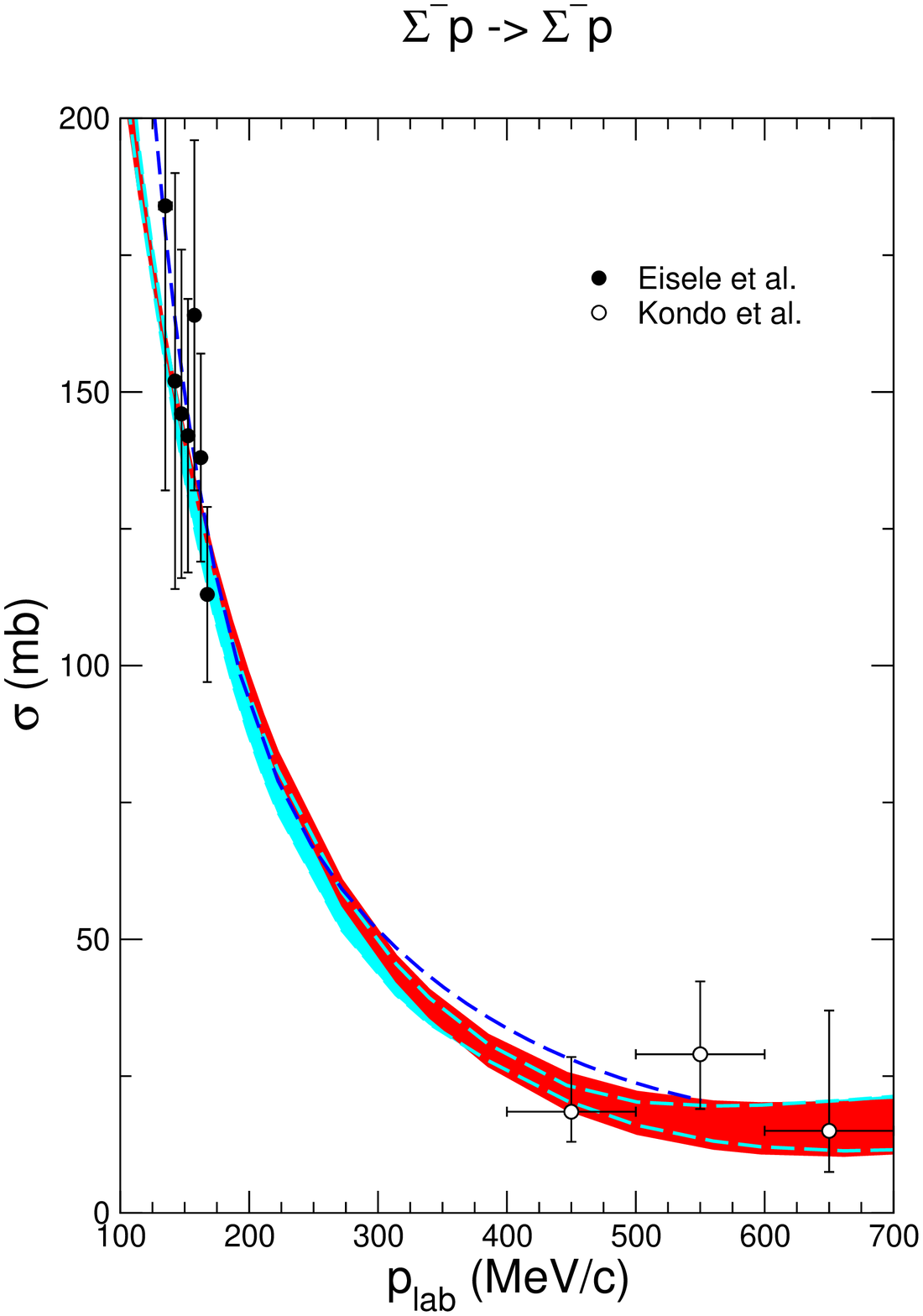}
\includegraphics[height=100mm]{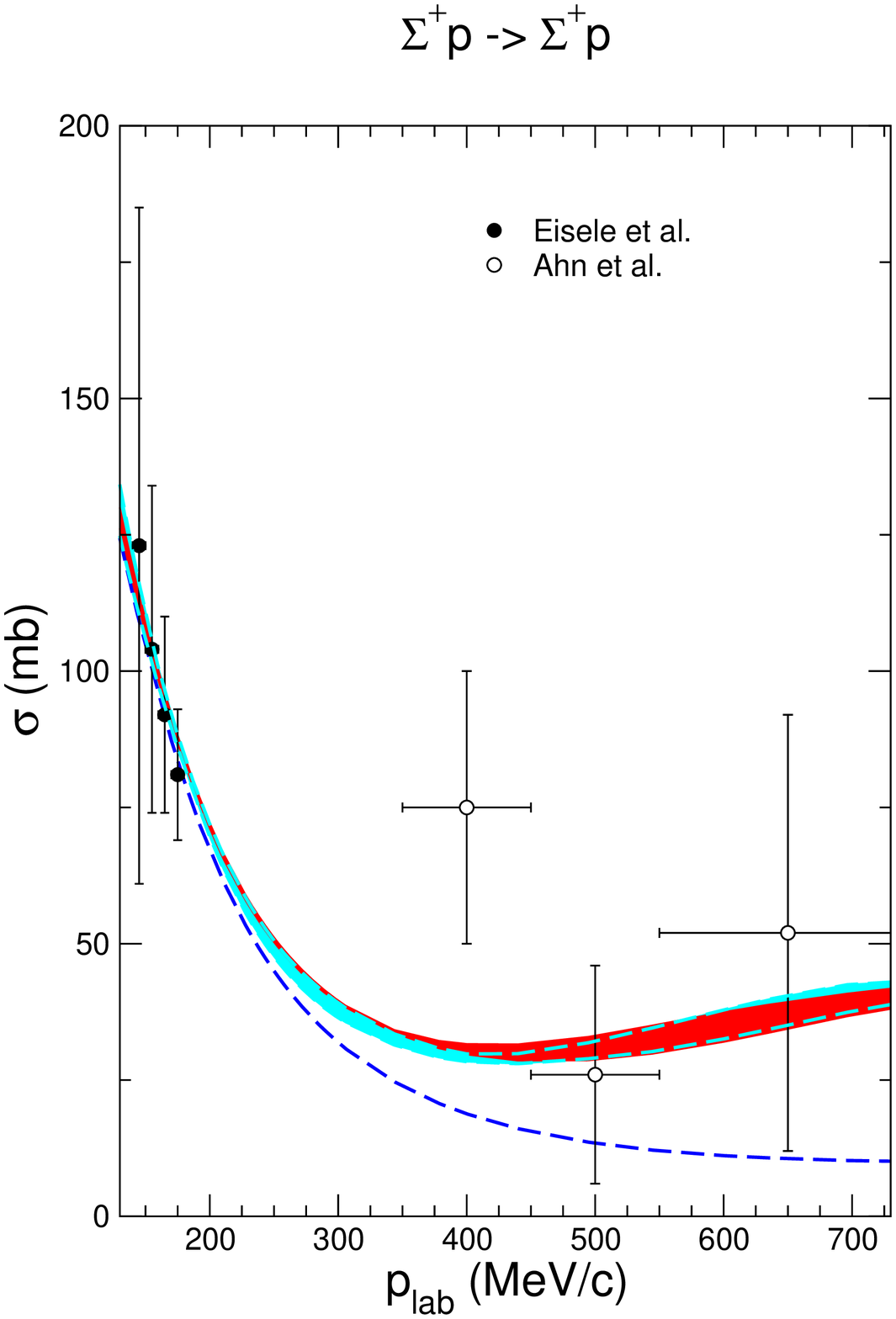}
\caption{Different $\Si$N and $\Si$N $\to$ $\La$N cross 
sections for higher energy. 
Same description of curves as in Fig.~\ref{fig:C1}.
The experimental cross sections
are taken from Refs.~\cite{Eng66,Ste70}
($\Sigma^-p \to \Lambda n$, $\Sigma^-p \to \Sigma^0 n$) 
and \cite{Eis71,Kon00} ($\Sigma^-p \to \Sigma^- p$), 
and \cite{Eis71,Ahn05} ($\Si^+ p \to \Si^+p$).
Note that those data at higher energy are not included in the fit. 
}
\label{fig:C3}
\end{center}
\end{figure*}

\begin{figure*}
\begin{center}
\includegraphics[height=100mm]{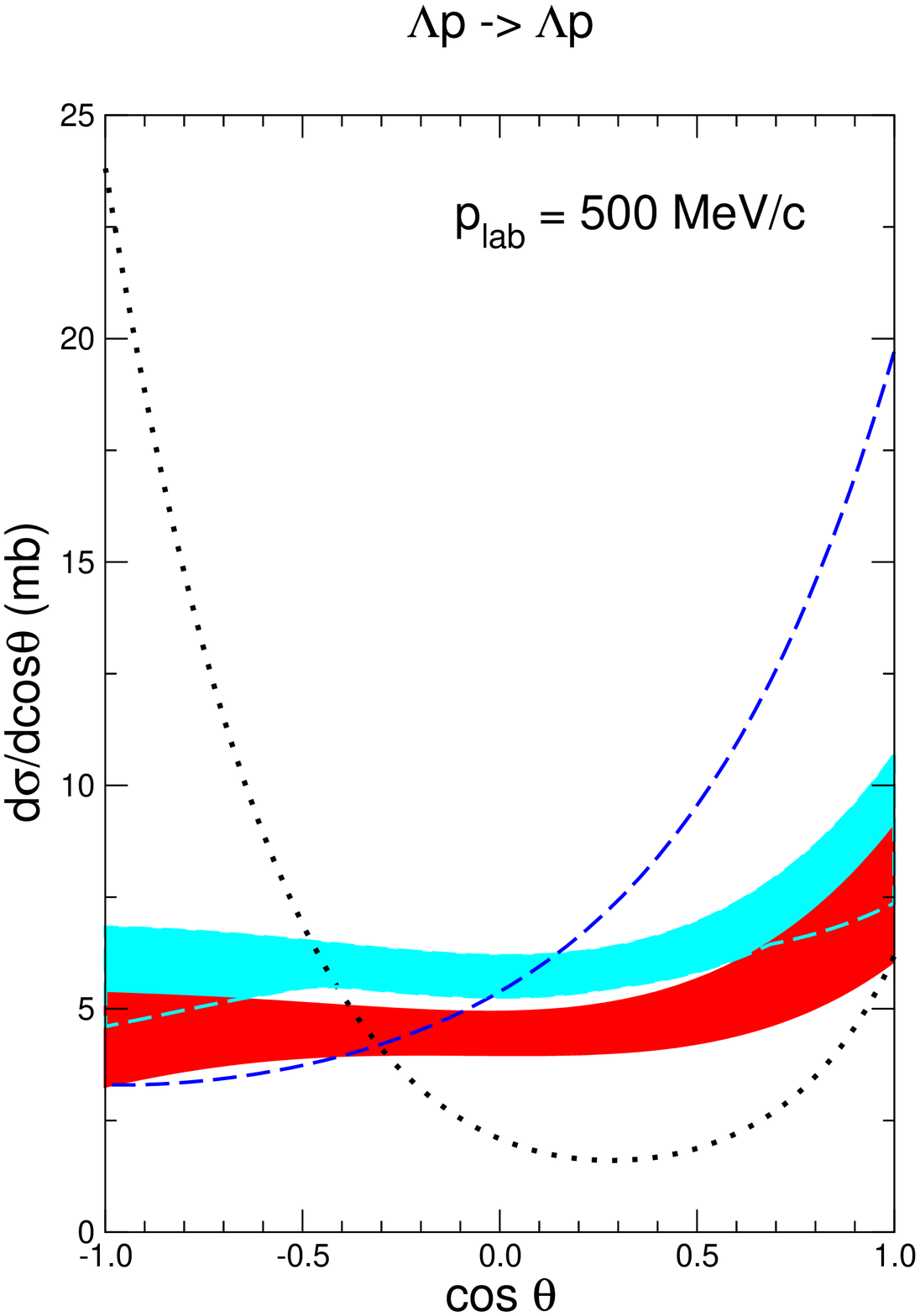}
\includegraphics[height=100mm]{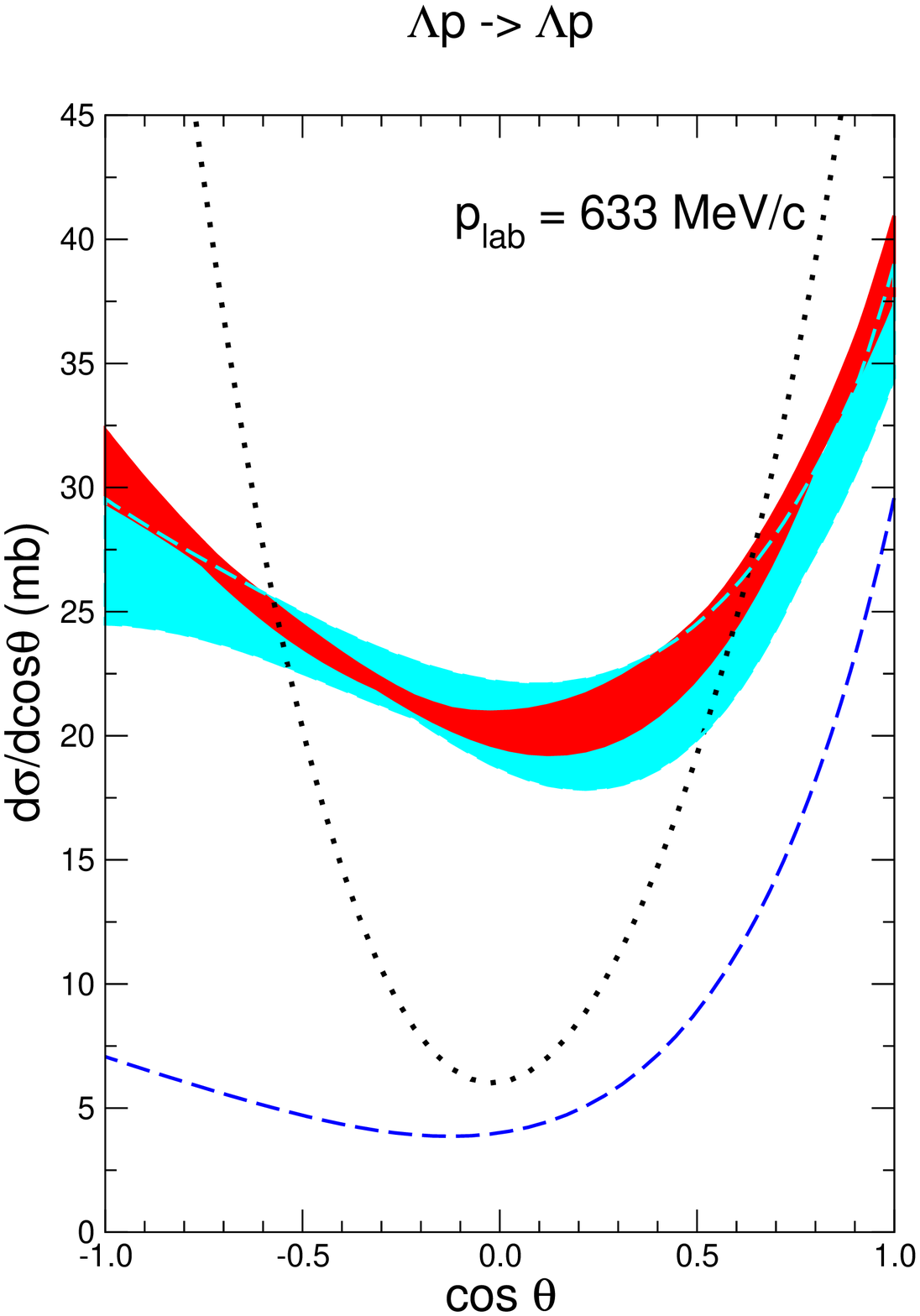}
\caption{Differential cross section for $\La p$ scattering at $500$ MeV/c 
and at $633$ MeV/c.
Same description of curves as in Fig.~\ref{fig:C1}.
} 
\label{fig:DN0}
\end{center}
\end{figure*}

\begin{figure*}
\begin{center}
\includegraphics[height=100mm]{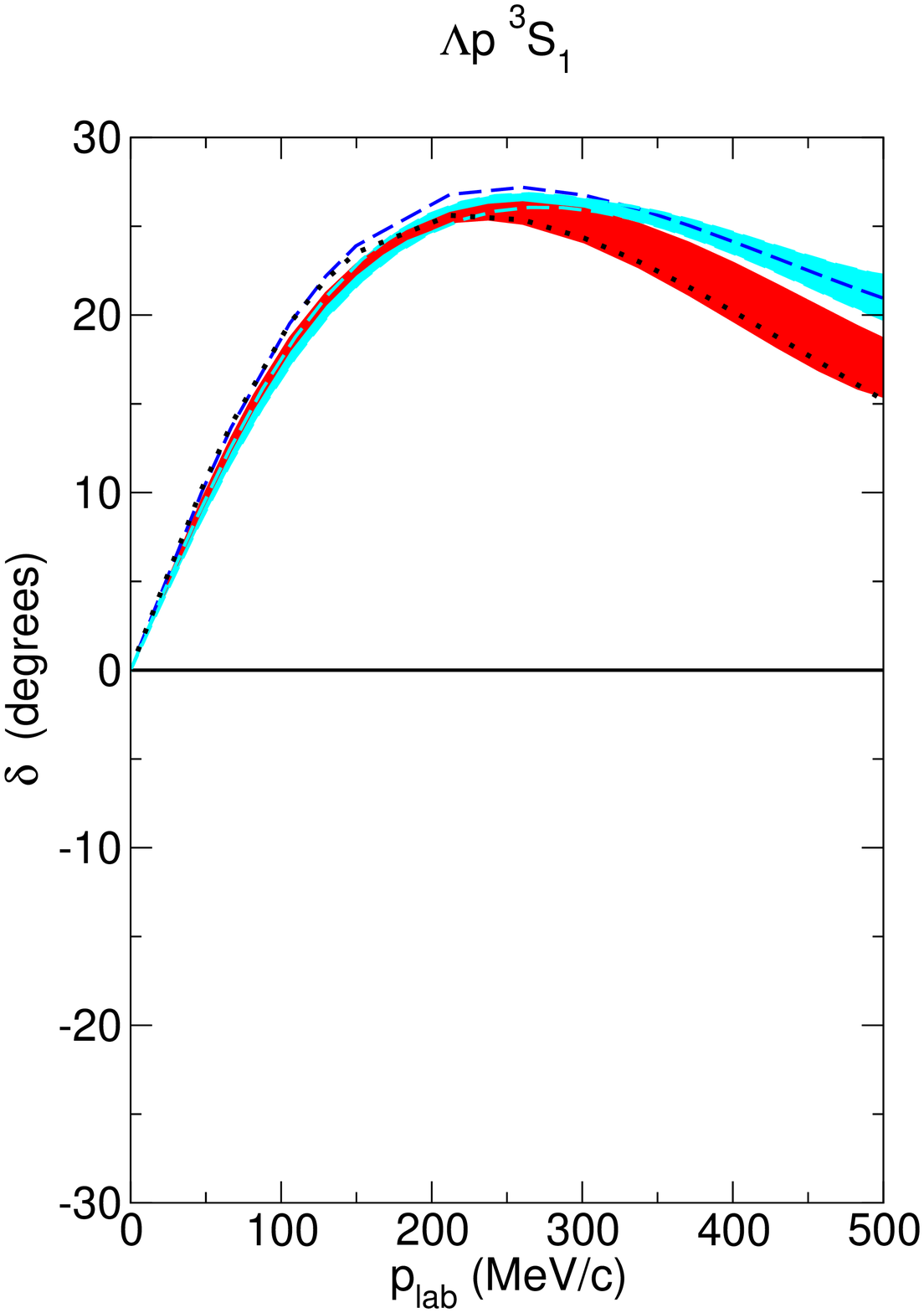}
\includegraphics[height=100mm]{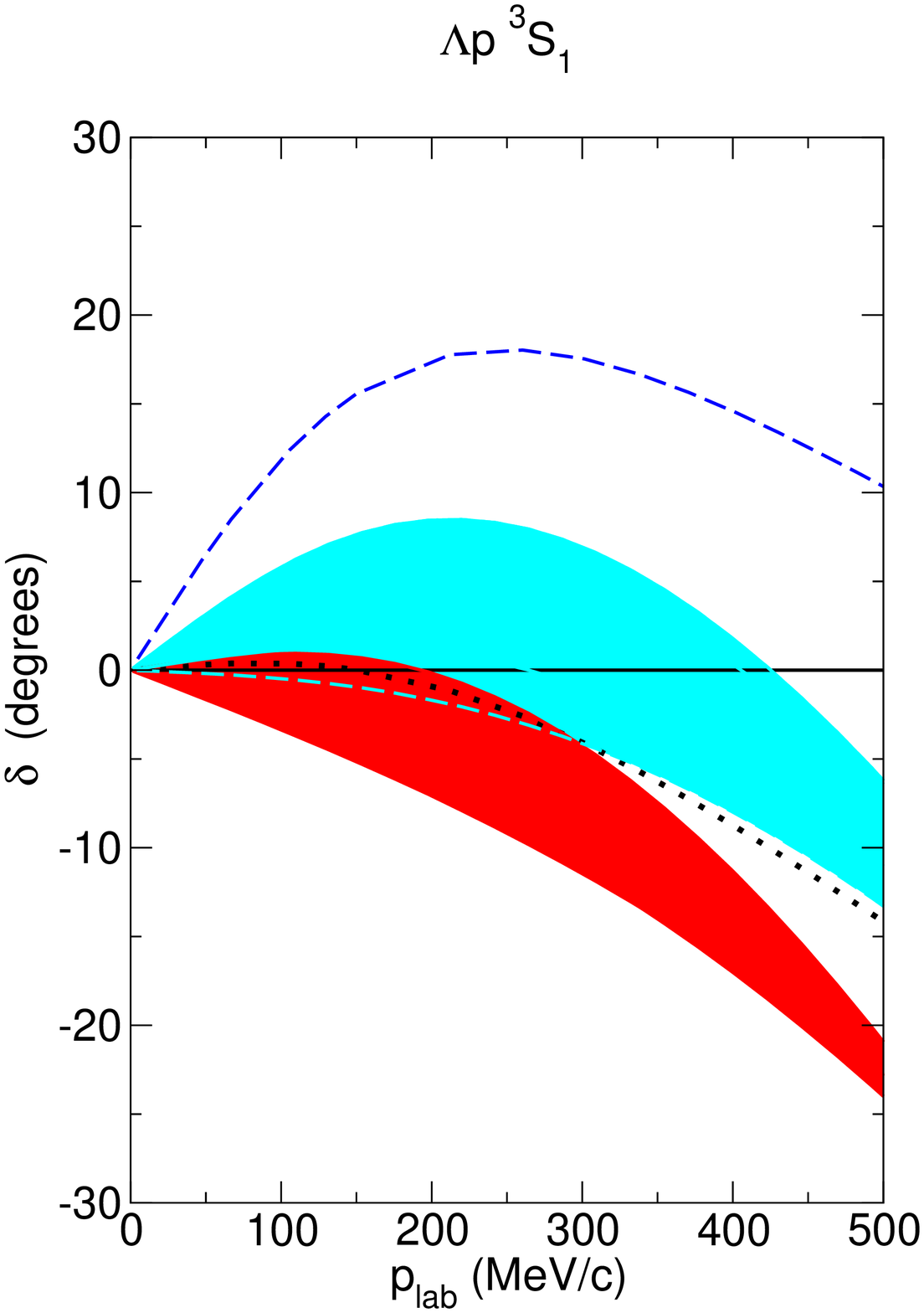}
\caption{$^3S_1$ $\La N$ phase shift with (left) and without (right) $\Si N$ coupling.
Same description of curves as in Fig.~\ref{fig:C1}.
}
\label{fig:Ph}
\end{center}
\end{figure*}

\begin{figure*}
\begin{center}
\includegraphics[height=100mm]{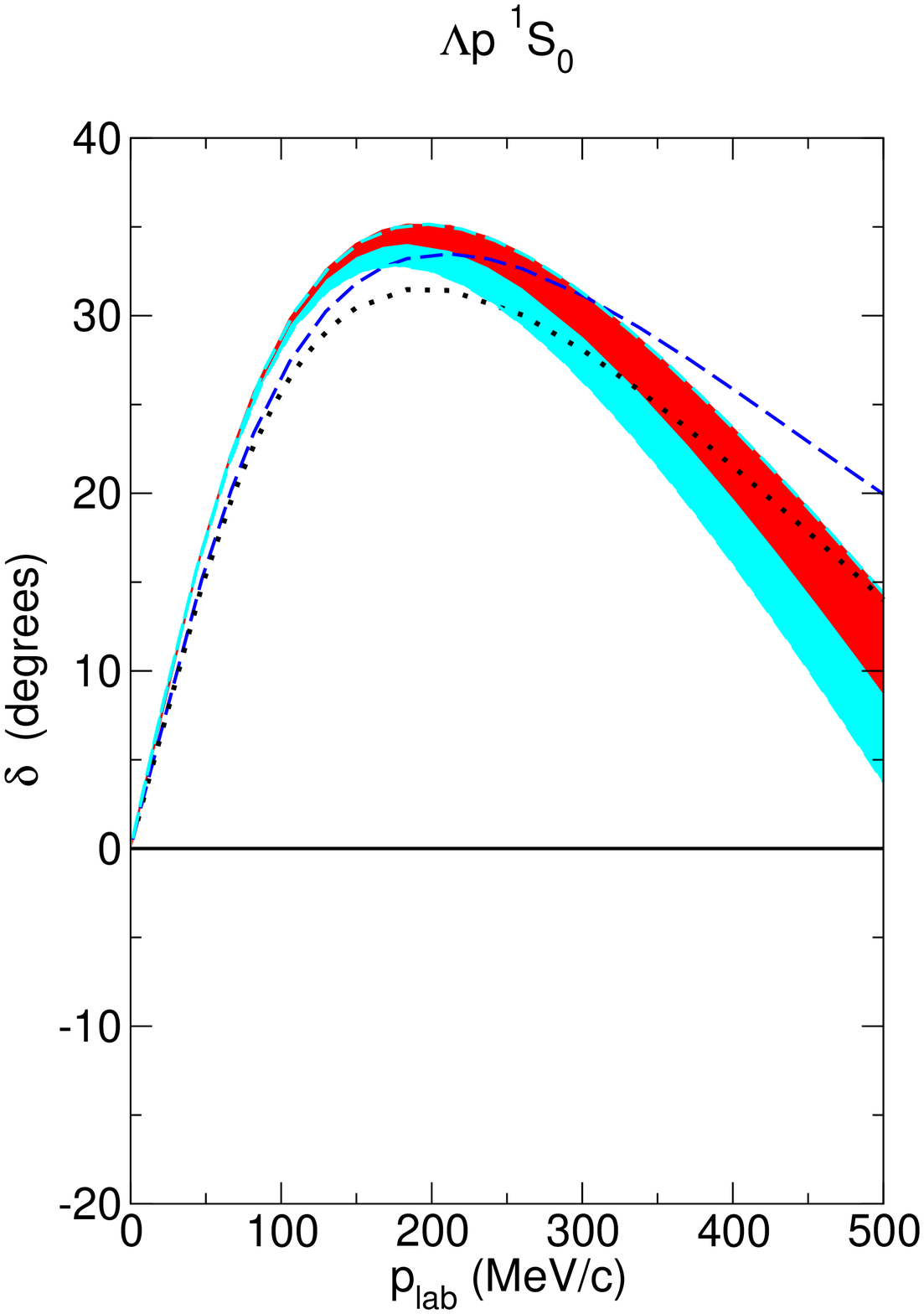}
\includegraphics[height=100mm]{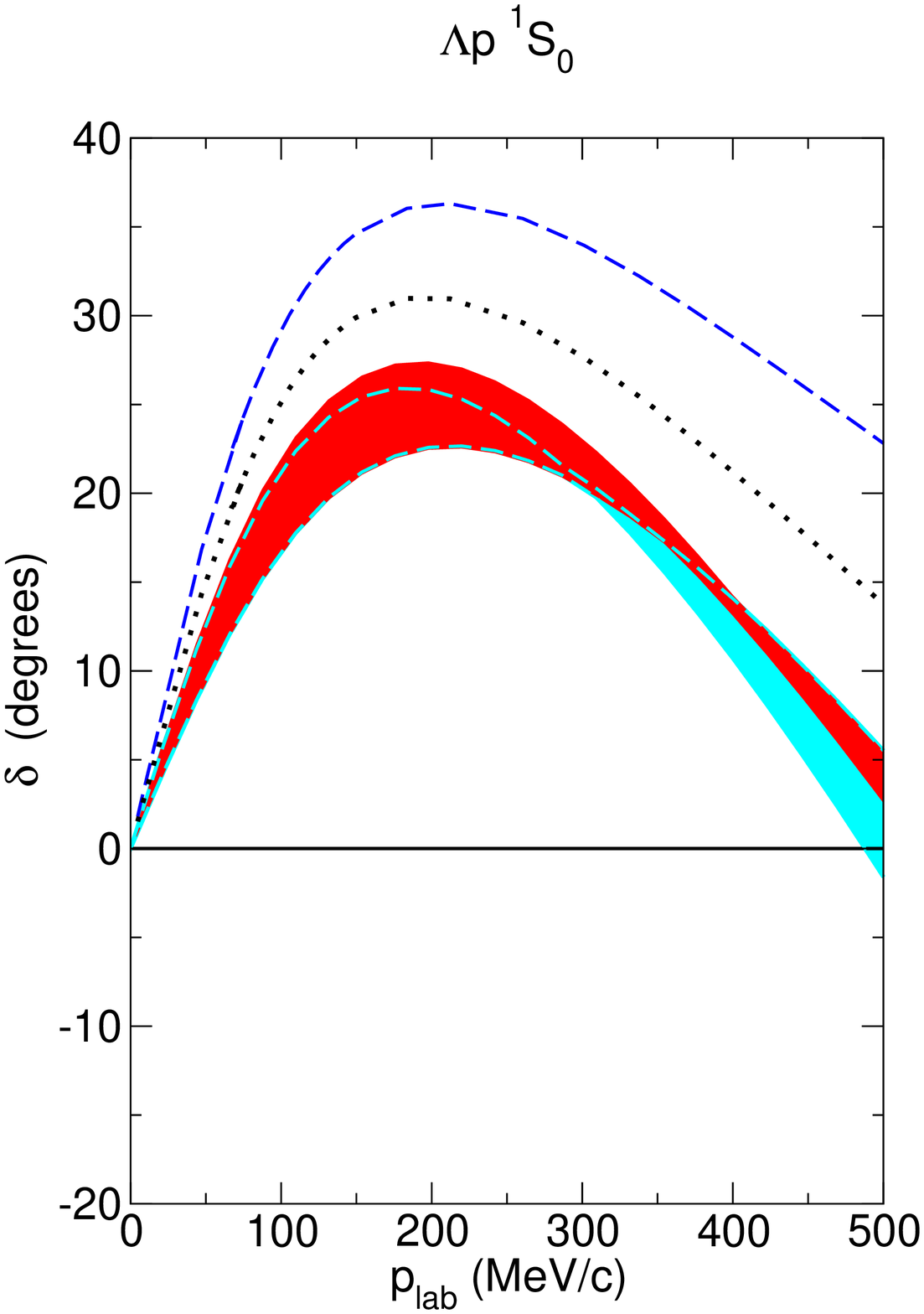}
\caption{$^1S_0$ $\La N$ phase shift with (left) and 
without (right) $\Si N$ coupling.
Same description of curves as in Fig.~\ref{fig:C1}.
}
\label{fig:Ph1s0}
\end{center}
\end{figure*}

\begin{figure*}[t]
\begin{center}
\includegraphics[height=100mm]{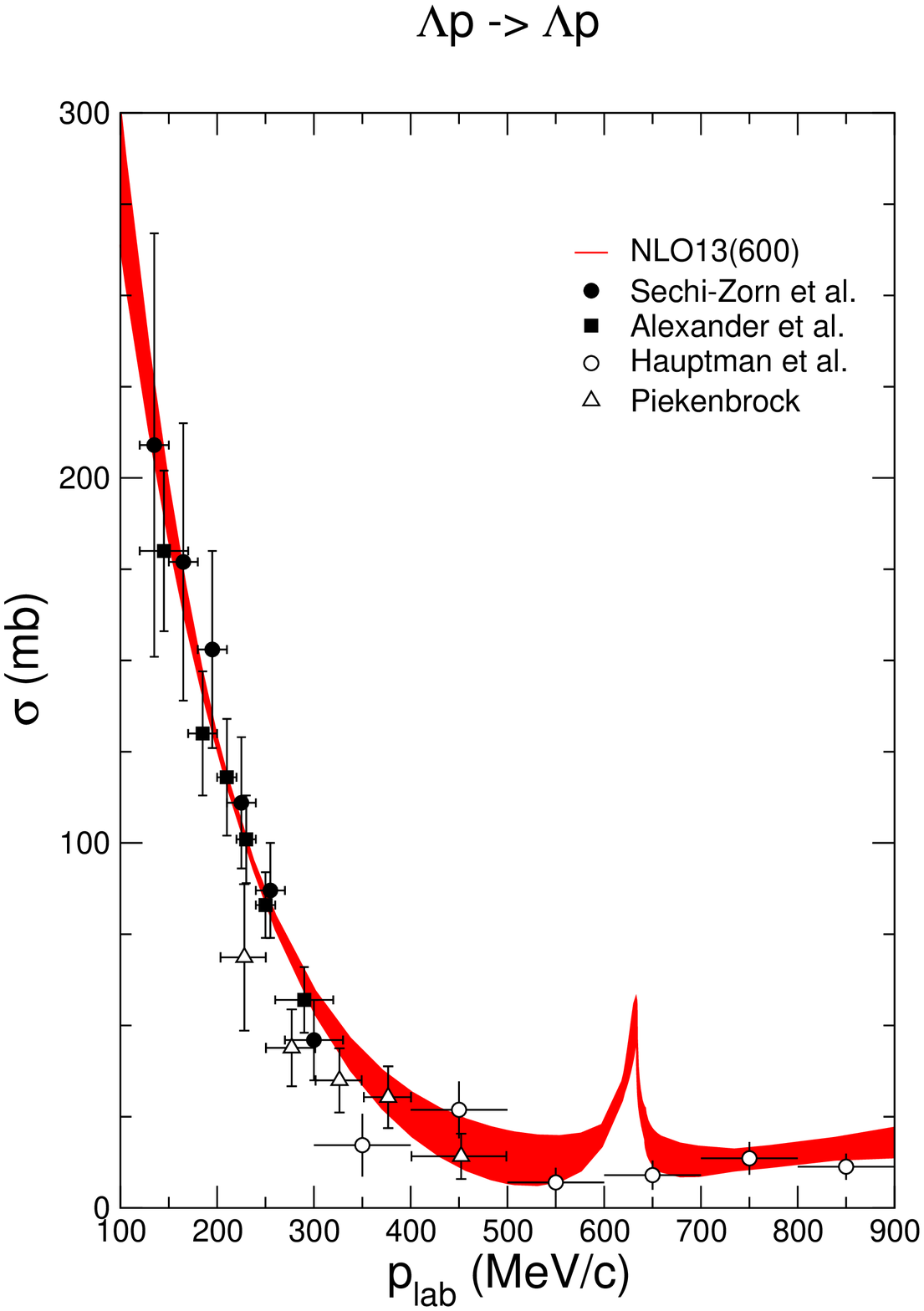}
\includegraphics[height=100mm]{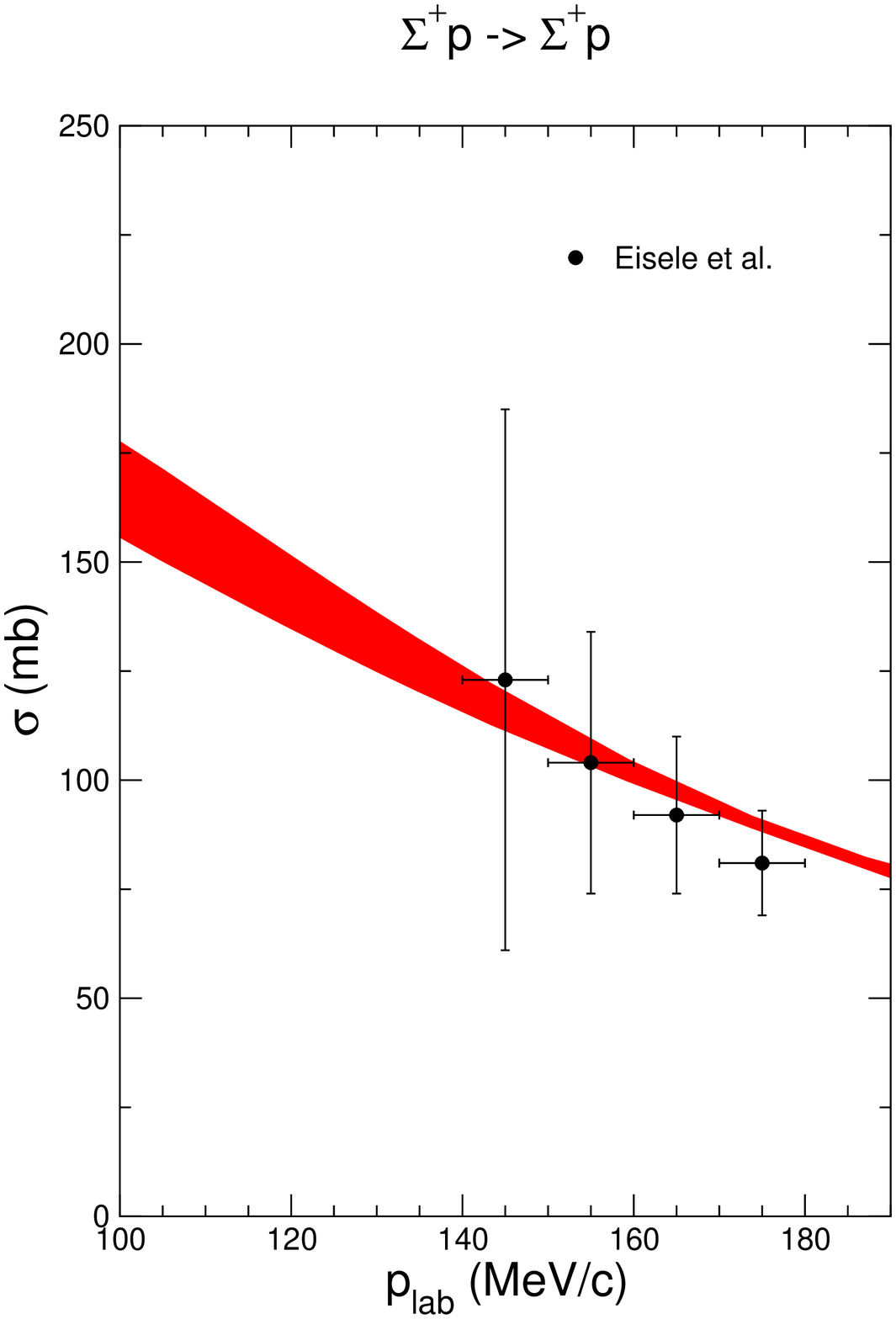}
\caption{
Uncertainty estimate for the $YN$ interaction in the
$\La p$ and $\Si^+ p$ channels employing the method suggested
in Ref.~\cite{Epelbaum:2015}. As basis the LO \cite{Polinder:2006} and NLO19 interactions
with cutoff $\Lambda=600$ MeV are used. We only show the NLO result and its uncertainty. 
}
\label{fig:E1}
\end{center}
\end{figure*}

\begin{figure*}
\begin{center}
\includegraphics[height=100mm]{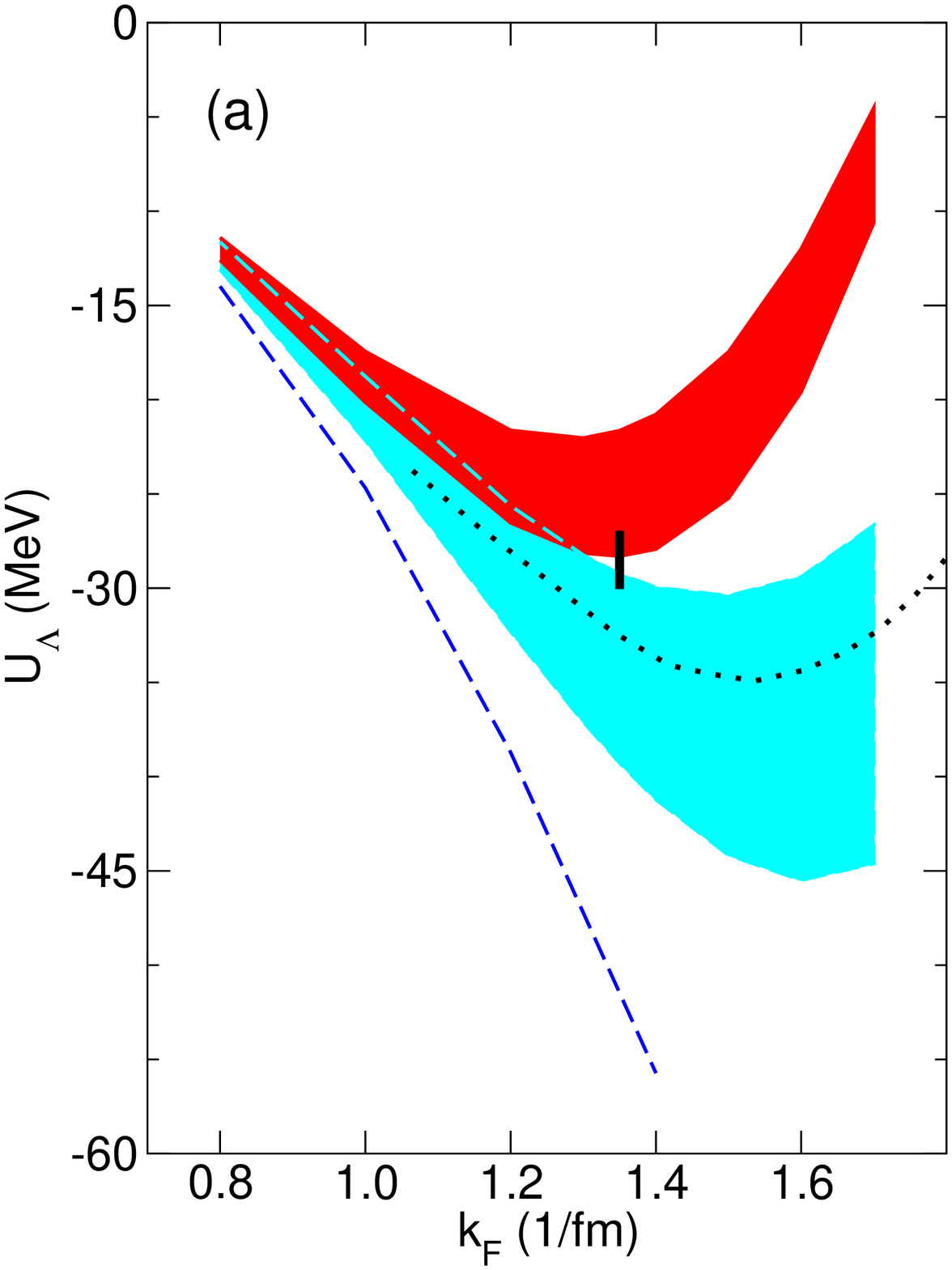}
\includegraphics[height=100mm]{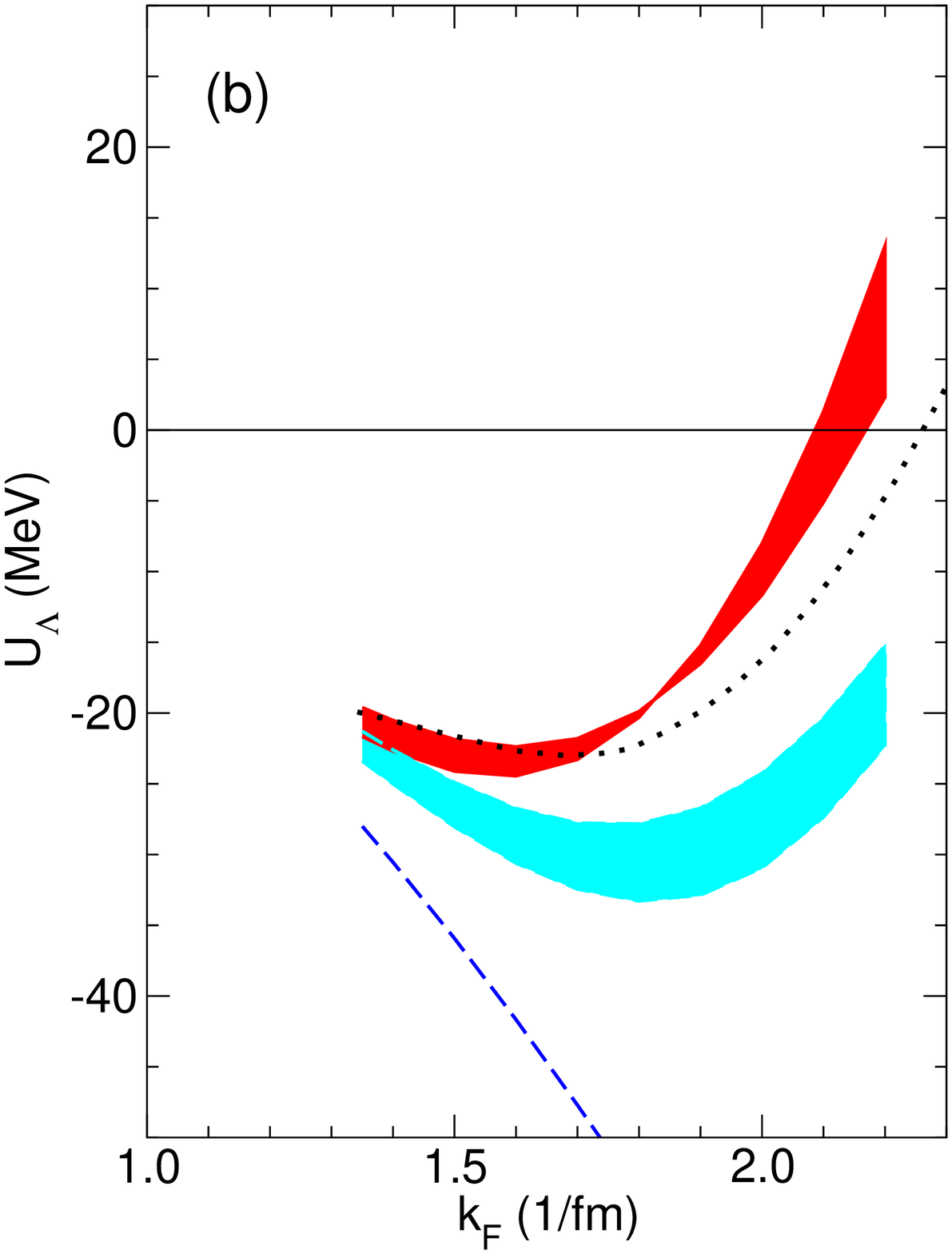}
\caption{The $\La$ single-particle potential $U_\La(p_\La = 0)$ as a function of the Fermi momentum $k_F$ in symmetric nuclear matter (a) and in neutron matter (b).
Same description of curves as in Fig.~\ref{fig:C1}.
The dotted curve is the result of the Nijmegen NSC97f 
potential \cite{Rijken:1999}, taken from Ref.~\cite{Yamamoto:2000}.
The vertical bar indicates the ''empirical'' value 
\cite{Gal:2016}. 
}
\label{fig:NML}
\end{center}
\end{figure*}

Obviously, the $\Lambda p$ cross sections produced by the two interactions 
are practically identical over the whole considered momentum range and hard
to distinguish in the plot. Visible differences occur only at higher momenta
near to the $\Si N$ thresholds where the NLO19 interaction 
predicts somewhat larger cross sections. 
There is also a more noticeable dependence 
of the results on the cutoff in the region below and above the thresholds. 
The latter is not too surprising because some LECs are 
fixed from $NN$ in the NLO19 interaction so that there is less flexibility to absorb the regulator 
dependence than in NLO13. Consequently, in general, a somewhat stronger variation 
of the cross sections with the cutoff has to be expected. 

Results for the various $\Si N$ channels are summarized in Fig.~\ref{fig:C2}.
Also here, there is practically no difference between the results of the
NLO13 and NLO19 interactions, except may be for the already mentioned
slightly increased dependence on the cutoff in case of the latter. 
Even at higher energies, the cross sections for the two interactions are
difficult to distinguish, see Fig.~\ref{fig:C3}. Note that these results
have to be considered as genuine predictions because none of the $\Si N$
data at momenta above $170$ MeV/c have been included  in the fitting procedure. 
The similarity of the predictions is particularly surprising in view of the
mentioned correlations between the LO and NLO LECs. One would have expected
that they are broken at higher momenta because the different values for the
LO and NLO LECs in the interactions NLO13 and NLO19 should yield a 
different energy dependence, at least when a larger energy region 
is considered. 

The predictions for differential cross sections, for \\  $\Si^-p\to \Si^-p$, 
$\Si^-p \to \La n$, and $\Si^+p \to \Si^+p$, at the few momenta were
data are available \cite{Haidenbauer:2013} remain basically unchanged  
and, therefore, we refrain from showing them. 
Instead, for illustrative purposes, we present results for the 
$\La p$ 
differential cross section at two selected laboratory momenta,
see Fig.~\ref{fig:DN0}. The momenta correspond to 
$T_{lab} = 107$ and $167.3$~MeV,
respectively, where the latter is just below the $\Si^+ n$ threshold. 
Again, the variations with the cutoff aside, there is 
hardly any difference between the predictions of the NLO13 and 
NLO19 interactions. One can see that for both potentials, at 
the lower energy, the
cross section is dominated by the $S$-waves whereas, at the 
$\Si^+ n$ threshold, there is a pronounced angular dependence 
that is actually induced by an
interference of the $^3S_1$-$^3D_1$ with the $^3P_2$ partial wave.
More striking are the differences to the predictions by 
the phenomenological potentials. 
In case of the J\"ulich '04 potential \cite{Haidenbauer:2005},
there is already a stronger angular dependence at the lower energy, indicating
a sizable contribution from $P$-waves. On the other hand, in the NSC97f
potential \cite{Rijken:1999}, there is a large contribution from the $^3D_1$ 
which is most obvious from the result at the $\Si^+ n$ threshold.
Evidently, experimental information would be very valuable 
here, but is, of course, rather difficult to obtain. 

Finally, the low-energy parameters for the $\Si N$ channels 
can be found in Table~\ref{tab:R0}, too. 
Besides the $\Si^+ p$ effective range parameters which include
the distortion from the Coulomb interaction, we list also the
$\Si N$ scattering length for the isospin $I=1/2$ channel calculated with an isospin-averaged $\Si$ mass. 
Also here, the variations in the predictions by the NLO13 and NLO19 potentials are small,
especially in case of the $^1S_0$ partial wave. 
Only in the $^3S_1$ partial wave with $I=1/2$, there is a more sizable difference, at least for the
lower cutoff values.  Here, the magnitude of the real and imaginary parts are 
noticeably different. There are also differences to the predictions of the phenomenological potentials.

\begin{figure}
\begin{center}
\includegraphics[height=100mm]{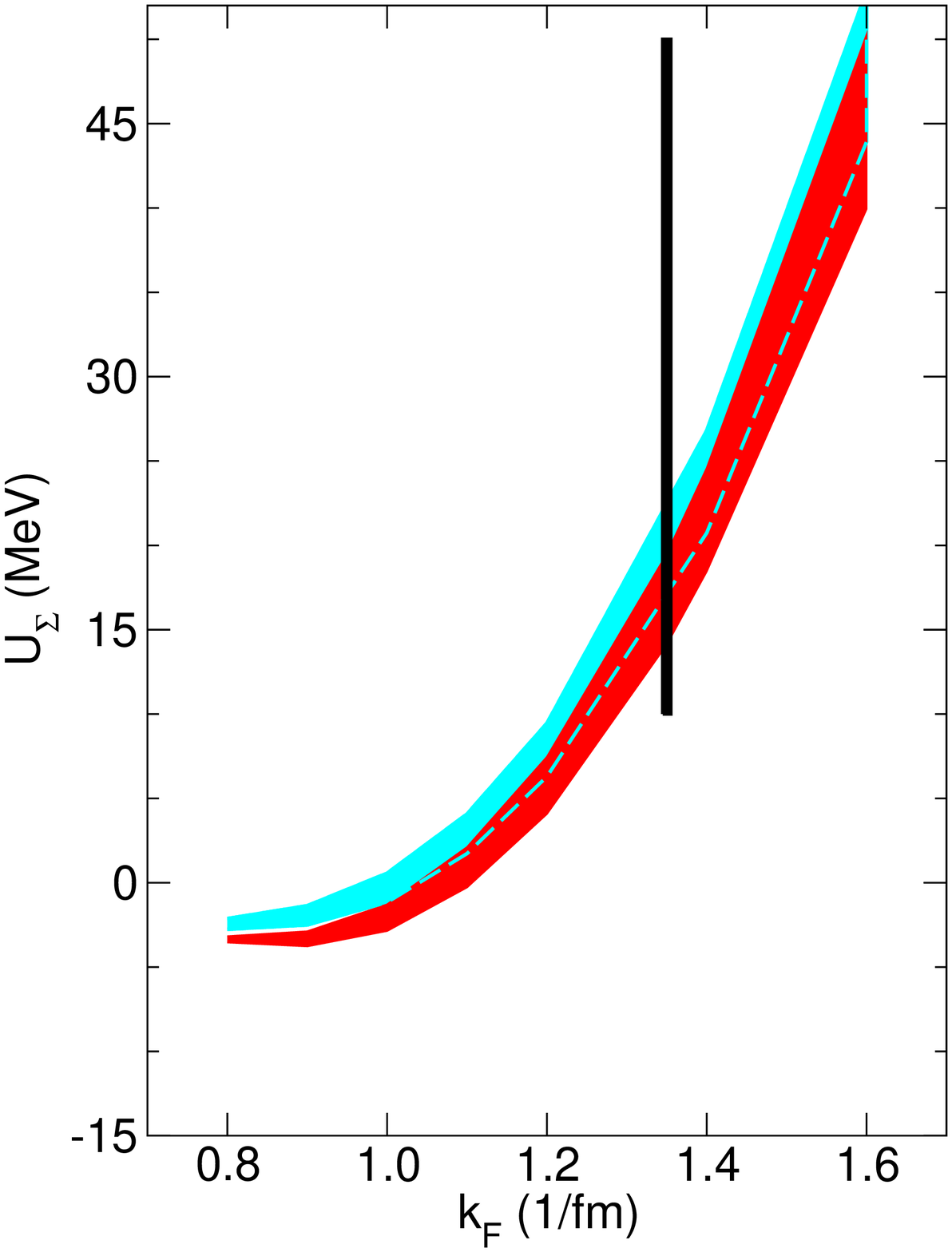}
\caption{The $\Si$ single-particle potential $U_\Si(p_\Si = 0)$
as a function of the Fermi momentum $k_F$ in symmetric nuclear
matter. Same description of curves as in Fig.~\ref{fig:C1}.
The vertical bar indicates the ''empirical'' value \cite{Gal:2016}.
}
\label{fig:NMS}
\end{center}
\end{figure}

This brings us to the question, whether there is any significant difference between the
$\La N$ and $\Si N$ scattering results  of the NLO13 and NLO19 potentials. And the simple
answer is that there is none, at least not in terms of observable quantities. 
That said, the just discussed $I=1/2$ $\Si N$ scattering 
length gives us a clue that there is a subtle difference and 
it concerns the strength of the $\La N$-$\Si N$ transition potential.
The simplest way to see that is to perform an ``academic'' calculation. It consists in simply switching
off the coupling potential between the two channels \cite{Haidenbauer:2017}. 
The outcome of such an exercise for the $\La N$ $^3S_1$ phase shift 
is presented in Fig.~\ref{fig:Ph}. The results on the left side are for
the full (coupled-channel) calculation and it is obvious that the phase 
shifts for the NLO13 and NLO19 potentials lie basically on top of each other,
at least up to momenta of $p_{lab}\approx 400$~MeV/c. On the right hand
side are the results without channel coupling. Here, one can see that
NLO13 (i.e. the $\La N$ potential alone) leads to mostly 
negative phase  shifts that are a sign for a repulsive potential, 
the J\"ulich '04 potential leads to a positive (attractive) phase, and the NLO19 potential
yields results somewhat in between. While such differences are not visible in two-body observables,
once evaluated for the full (coupled-channel) potential  (cf. the results presented above), 
they do have an influence in applications to few- and many-body 
systems, to be discussed in the next subsections, even when
the full $\La N$-$\Si N$ potential is used. 

For completeness, we show also the corresponding
results for the $^1S_0$ partial wave, cf. Fig.~\ref{fig:Ph1s0}. 
Here, NLO13 and NLO19 behave alike. In both cases, there is a 
moderate reduction of the attraction when the coupling to
$\Si N$ is switched off. Differences occur only for the two
phenomenological potentials where the result for the $^1S_0$ phase
shift remains practically unchanged (NSC97f) or even increases
(J\"ulich '04) without $\Si N$ coupling.

Note that the cutoff dependence increases when the coupling is switched off. This happens because we use the (diagonal)
$\La N$ potential as established in the full coupled-channel
calculation. No re-adjustment of the contact terms is done and,
thus, there is no proper absorption of the regulator dependence
in this ''academic'' calculation. 

Finally, for illustration, we present an estimate for
the theoretical uncertainty following the method 
proposed in Ref.~\cite{Epelbaum:2015}. 
In Fig.~\ref{fig:E1}, selected results for the NLO19 
potential for the cutoff $\Lambda=600$~MeV are shown.
This value is also used as breakdown scale \cite{Epelbaum:2015}.
For this estimate, the difference of the LO results \cite{Polinder:2006}
and the NLO result is used for an estimated of the uncertainty. 
As stated already in Sect.~\ref{sec:Formalism}, at the NLO
level, it is premature to address the question of convergence.
For this endeavor, more orders are required to avoid that accidentally close 
results lead to an under estimation of the uncertainty. For the $YN$ interaction, 
this uncertainty estimate is especially difficult since 
the data is not sufficient to unambiguously determine all LECs. 
For this reason, it is also not useful to quantify the uncertainty 
of phase shifts of individual partial waves in this manner. 
Nonetheless, we want to emphasize that the estimated
uncertainty appears sensible and also plausible.  
In particular, it encases the variations due to the regulator
dependence and, thus, is consistent with the expectation
that cutoff variations provide a lower bound for the
theoretical uncertainty. 
For details of the method and a thorough discussion of
the underlying concept, we refer the reader to \cite{Binder:2015mbz}.

\subsection{{\boldmath$\La $} and {\boldmath$\Si $} in nuclear matter}
\label{sec:Matter}

Let us now compare the in-medium properties of the $YN$ interactions NLO13 and NLO19.
Table~\ref{tab:R3} summarizes the values for the $\Lambda$ and $\Sigma$ potential 
depths, $U_\Lambda (p_\Lambda = 0)$ and $U_\Sigma (p_\Sigma = 0)$, evaluated at the 
saturation point of nuclear matter, i.e. for $k_F=1.35$ fm$^{-1}$. Note that the results for
NLO13 slightly differ  from those given in \cite{Haidenbauer:2015} because a
different and more up to date nucleon s.p. potential is used,  see Sect.~\ref{sec:MatterF}. 
Corresponding results obtained for the J\"ulich'04 
meson-exchange potential \cite{Haidenbauer:2005} and the 
Nijmegen NSC97f potential \cite{Rijken:1999} are also included. 
The dependence of the hyperon potential depths on the Fermi momentum is
displayed in Figs.~\ref{fig:NML} and \ref{fig:NMS}. 

It is quite obvious from Fig.~\ref{fig:NML} that the EFT potential NLO19 is much more
attractive in the medium than  NLO13. The difference is primarily due to the contribution 
of the $^3S_1$-$^3D_1$ partial wave which is enhanced by  more or less a factor $2$ for the
new interaction, see Table~\ref{tab:R3}. Actually, the density dependence predicted 
by NLO19 is similar to the one of the NSC97f potential, cf. 
the dotted line in Fig.~\ref{fig:NML}. It is instructive to 
compare the figure for $U_\La$ with the one for the $^3S_1$ 
phase shifts with the $\La N$-$\Si N$ coupling switched off
(right-hand side of Fig.~\ref{fig:Ph}). One can easily see that 
a stronger contribution of the conversion, leading to most changes in 
Fig.~\ref{fig:Ph}, at the same time, leads to a smaller single-particle 
potential. NSC97f is still in between NLO13 and NLO19 although the phase 
shift in the figure is not in complete agreement with NLO19 anymore. 
Nevertheless, the outcome for the single-particle potential of the $\Lambda$ seems to be strongly 
influenced by the strength of the $\La N$-$\Si N$ coupling potential
\cite{Nogami:1970,Bodmer:1971,Dabrowski:1973}.  
For the NLO13 interaction, the influence of the strength of the transition potential on
the in-medium properties of the $\La$  was already discussed in detail by some of us in 
Ref.~\cite{Haidenbauer:2017} and subsequently by Kohno \cite{Kohno:2018}. 

Fig.~\ref{fig:NML} reveals also that there is a sizable
and certainly unsettling cutoff dependence of the predictions. 
However, this is not too surprising given that a likewise  
strong regulator dependence has already been detected 
in corresponding studies of
nuclear matter properties in the $NN$ sector within chiral 
EFT \cite{Coraggio:2013,Sammarruca:2015,Hu:2016nkw}.
Since the Pauli operator in Eq.~(\ref{Eq:G1}) suppresses
the contributions from low momenta, the $G$-matrix results
are more sensitive to higher momenta and, thus, to intermediate
and short-distance physics \cite{Hu:2016nkw}. In the $NN$ case,
indications for a convergence and a reduced regulator 
dependence were only found after going to much higher order
- N$^3$LO in Refs.~\cite{Coraggio:2013,Sammarruca:2015} and N$^4$LO in \cite{Hu:2016nkw}
- and after including three-body forces.  Indeed, as argued in Ref.~\cite{Hu:2016nkw},
the cutoff dependence could allow one to draw indirect 
conclusions on the size of such many body forces.  

For completeness, we also show results for a $\Lambda$ in neutron matter
(right-hand side of Fig.~\ref{fig:NML}). Also in this case the $\La$ s.p. potential
predicted by NLO19 is much more attractive than the one by NLO13. Though there is a trend
to repulsion with increasing density, similar to NLO13 and the
NSC97f potential, it is clear that the actual change of sign 
will take place at significantly higher densities. 

Investigations of (finite) $\La$ hypernuclei utilizing the
EFT interactions are presently on the way \cite{Vidana:2019},
based on the formalism described in Ref.~\cite{Vidana:2017}. For 
even lighter hypernuclei, the interactions are also currently studied 
\cite{HLe}.

\begin{table*}[ht]
\caption{$\La$ and $\Si$ single-particle potentials $U_Y(p_Y=0)$ (in MeV)
at nuclear matter saturation density ($k_F = 1.35$ fm$^{-1}$).
The contributions from the $S$ waves and the total result including all partial waves up to $J=5$ are given.
  }
 \label{tab:R3}
\vskip 0.1cm
\renewcommand{\arraystretch}{1.4}
\begin{center}
\begin{tabular}{|c|rrrr|rrrr|rr|}
\hline
& \multicolumn{4}{|c|}{NLO13} & \multicolumn{4}{|c|}{NLO19} & \ J\"ulich '04 \ & \ NSC97f \ \\
\hline
${\Lambda}$ [MeV] & 500 & 550 & 600 & 650 &  500 & 550 & 600 & 650 &  & \\
\hline
\hline
\multicolumn{11}{|c|}{$U_\Lambda(0)$} \\
\hline
$^1S_0$           & $-15.3$ & $-13.7$ & $-12.3$ & $-11.3$ & $-12.5$ & $-11.6$ & $-11.2$ & $-11.1$ 
& $-10.2$ & $-14.6$ \\
$^3S_1$-$^3D_1$   & $-14.6$ & $-11.4$ & $-10.8$ & $-12.5$ & $-28.0$ & $-27.2$ & $-22.8$ & $-19.7$ 
& $-36.3$ & $-23.1$ \\
total             & $-28.3$ & $-23.5$ & $-21.6$ & $-22.3$ & $-39.3$ & $-37.1$ & $-32.6$ & $-29.2$ 
& $-51.2$ & $-32.4$ \\
\hline
\hline
\multicolumn{11}{|c|}{$U_\Sigma(0)$} \\
\hline
$^1S_0$ (1/2)         &$  6.9$ & $  6.4$ & $  5.0$ & $  4.4$ & $  6.7$ & $  6.3$ & $  5.0$ & $  4.4$ 
& $  4.2$ & $ 15.0$ \\
$^1S_0$ (3/2)         &$-11.4$ & $-10.7$ & $-10.1$ & $ -9.5$ & $-10.8$ & $-10.4$ & $- 9.9$ & $ -9.5$ 
& $-12.0$ & $-12.6$ \\
$^3S_1$-$^3D_1$ (1/2) &$-21.7$ & $-22.9$ & $-22.7$ & $-21.7$ & $-18.0$ & $-17.6$ & $-20.0$ & $-20.3$ 
& $-15.0$ & $- 8.8$ \\
$^3S_1$-$^3D_1$ (3/2) &$ 40.0$ & $ 44.8$ & $ 43.6$ & $ 40.0$ & $ 41.0$ & $ 38.0$ & $ 40.2$ & $ 38.7$ 
& $ 11.7$ & $- 6.4$ \\
total                 &$ 16.7$ & $ 19.4$ & $ 17.1$ & $ 14.1$ & $ 21.6$ & $ 18.4$ & $ 16.6$ & $ 14.1$ 
& $-22.2$ & $-16.1$ \\
\hline
\hline
\end{tabular}
\end{center}
\renewcommand{\arraystretch}{1.0}
\end{table*}

Results for the $\Si$ s.p. potential in symmetric nuclear matter are presented 
in Fig.~\ref{fig:NMS}. It is predicted to be repulsive by NLO13 as well as by NLO19, 
in agreement with evidence from the analysis of level shifts and
widths of $\Sigma^-$ atoms and from measurements of $(\pi^-,K^+)$ inclusive 
spectra related to $\Sigma^-$-formation in heavy nuclei \cite{Gal:2016}.
As discussed in detail in Ref.~\cite{Haidenbauer:2015}, a repulsive $\Sigma$ s.~p. 
potential is achieved because the EFT interactions in the $^3S_1$ partial 
wave of the $\Si^+ p$ channel (which provides the dominant contribution, cf. 
Table~\ref{tab:R3}) are repulsive. 
Note that a repulsive $^3S_1$ interaction is in accordance with results from 
lattice QCD calculations \cite{Beane:2012,Nemura:2018}.
The NLO19 interaction provides slightly more repulsion. But overall, with
regard to the $\Sigma$ in-medium properties, there is very little difference
to NLO13. This is also true on the level of the partial-wave contributions, as can be
seen by comparing the corresponding values in Table~\ref{tab:R3}.  
 
As exemplified by the predictions of the J\"ulich '04 and NSC97f potentials, 
typically phenomenological potentials fail to produce a repulsive $\Sigma$-nuclear 
potential, cf. the corresponding results in Table~\ref{tab:R3}. Because of 
that we refrain from showing the pertinent curves in Fig.~\ref{fig:NMS}. 

\begin{table*}
  \caption{Dependence of the separation energies of $^3_\Lambda {\rm H}$, $^4_\Lambda {\rm He(0^+)}$ and
    $^4_\Lambda {\rm He(1^+)}$ for NLO19(650) on the $NN$ interaction. 
    The $\Sigma$ probabilities are shown, too. 
    Energies are given in MeV, probabilities in \%. 
    The semilocal momentum-space (SMS) chiral $NN$ interaction 
    from Ref.~\cite{Reinert:2017} is employed. 
    }
    \label{tab:R4NN}
    \centering
    \renewcommand{\arraystretch}{1.4}
\begin{tabular}{|r|ccc|ccc|}
\hline
    $NN$ interaction     & $E_{\Lambda}(^3_\Lambda {\rm H})$  & $E_{\Lambda}(^4_\Lambda {\rm He(0^+)})$  & $E_{\Lambda}(^4_\Lambda {\rm He(1^+)})$   
                       & $P_{\Sigma}(^3_\Lambda {\rm H})$ & $P_{\Sigma}(^4_\Lambda {\rm He(0^+)})$  & $P_{\Sigma}(^4_\Lambda {\rm He(1^+)})$ \\
\hline
  SMS N$^4$LO(400)  &   0.099   &     1.556 &    0.921  &     0.223 &     1.533 &     1.527  \\
  SMS N$^4$LO(450)  &   0.097   &     1.542 &    0.916  &     0.222 &     1.526 &     1.522  \\
  SMS N$^4$LO(500)  &   0.093   &     1.509 &    0.894  &     0.218 &     1.509 &     1.506  \\
  SMS N$^4$LO(550)  &   0.089   &     1.472 &    0.870  &     0.213 &     1.490 &     1.486  \\
  \hline
\end{tabular}    
\renewcommand{\arraystretch}{1.0}
\end{table*}

\begin{table*}
  \caption{Dependence of the separation energies $E_\Lambda$ of $^3_\Lambda {\rm H}$,
    $^4_\Lambda {\rm He(0^+)}$ and $^4_\Lambda {\rm He(1^+)}$ on the $YN$ interaction.The $\Sigma$
    probabilities $P_\Sigma$ are also shown. For the chiral YN forces, the SMS 
    $NN$ interaction \cite{Reinert:2017} at order N$^4$LO+ 
    with cutoff of 450~MeV has been used. For J\"ulich'04 and NSC97f, the CD-Bonn interaction \cite{CDBonn}
    has been employed.     
    Energies are given in MeV, probabilities in \%. 
    Experimental values are taken from 
    Refs.~\cite{Davis:1992dt} ($^3_\Lambda {\rm H}$,  
    $^4_\Lambda {\rm He(0^+)}$) and 
    \cite{Yamamoto:2015} ($^4_\Lambda {\rm He(1^+)}$).
    }
    \label{tab:R4new}
    \centering
    \renewcommand{\arraystretch}{1.4}
\begin{tabular}{|r|ccc|ccc|}
\hline
      YN interaction & $E_{\Lambda}(^3_\Lambda {\rm H})$  & $E_{\Lambda}(^4_\Lambda {\rm He(0^+)})$  & $E_{\Lambda}(^4_\Lambda {\rm He(1^+)})$   
                       & $P_{\Sigma}(^3_\Lambda {\rm H})$ & $P_{\Sigma}(^4_\Lambda {\rm He(0^+)})$  & $P_{\Sigma}(^4_\Lambda {\rm He(1^+)})$\\
\hline
 NLO13(500) &   0.135   &    1.705 &   0.790 &     0.291 &     2.014 &     1.640 \\
 NLO13(550) &   0.097   &    1.503 &   0.586 &     0.273 &     2.108 &     1.556 \\
 NLO13(600) &   0.090   &    1.477 &   0.580 &     0.251 &     2.024 &     1.505 \\
 NLO13(650) &   0.087   &    1.490 &   0.615 &     0.232 &     1.870 &     1.397 \\
 \hline
 NLO19(500) &   0.100   &    1.643 &   1.226 &     0.168 &     1.120 &     1.261 \\
 NLO19(550) &   0.094   &    1.542 &   1.239 &     0.189 &     1.156 &     1.434 \\
 NLO19(600) &   0.091   &    1.462 &   1.055 &     0.208 &     1.368 &     1.676 \\
 NLO19(650) &   0.095   &    1.530 &   0.916 &     0.219 &     1.520 &     1.523 \\
 \hline 
 J\"ulich'04 &  0.046   &    1.704 &   2.312 &     0.181 &     0.782 &     0.895 \\
 NSC97f      &  0.099   &    1.832 &   0.575 &     0.190 &     1.798 &     1.078 \\
 \hline 
 Expt.       &  0.13(5) &  2.39(3) &   0.98(3) &      -- &        -- &       --  \\
 \hline
 \end{tabular}  
 \renewcommand{\arraystretch}{1.0}
\end{table*}

\subsection{Three- and four-body systems}
\label{sec:YNN}

In this section, we present results 
for the $^{\,3}_\La \rm H$ and $^{\,4}_\La \rm He$
binding energies based on the 
NLO13 and NLO19 $YN$ potentials and for the phenomenological 
Nijmegen (NSC97f) \cite{Rijken:1999} and
J\"ulich '04 \cite{Haidenbauer:2005} $YN$ interactions. 
We want to emphasize that the binding energies of the 
hypernuclei also depend on the employed $NN$ interaction 
and are affected by three-nucleon forces (3NFs). 
However, detailed calculations 
show that this dependence enters through the binding of the 
$3N$ core nucleus. It is therefore useful to present the results 
in terms of the difference of the core nucleus binding energy and 
the hypernuclear binding energy, the $\La$ separation energies,
which are denoted by $E_\La$ in the following. 
This fact is exemplified in Table~\ref{tab:R4NN} for the $YN$ 
interaction NLO19 with cutoff $\La=650$ MeV in 
combination with the high-order semilocal momentum-space 
regularized chiral $NN$ potential (SMS)
\cite{Reinert:2017} with different cutoffs. 
One can see that the $\La$ separation energy for 
$^{\,3}_\La \rm H$ varies only by $10$~keV. 
In case of $^{\,4}_\La \rm He$ the variations are in the
order of $80$ and $40$~keV for the $0^+$ and $1^+$
states, respectively. Similarly, small variations have
been found in calculations where phenomenological $NN$
potentials were employed \cite{Nogga:2013}. 
The addition of a 3NF  changes the binding energy by approximately $800$~keV
(depending on the chosen NN interaction) but the separation 
energy only by $20$-$50$~keV \cite{Nogga:2002}. In the following, we can therefore 
discuss the predictions for the separation energies 
independently from the $NN$ and $3N$ interactions.

In former studies of hyperonic few-body systems, 
the role of the spin-dependence of the $\Lambda N$ potential 
for the binding energies of s-shell hypernuclei has been 
discussed in terms of the appropriately averaged 
effective $\La N$ interaction \cite{Herndon:1967,Carlson:1991,Gibson:1994}.
We will do the same here. It is rather instructive 
and allows for a good qualitative understanding of 
the corresponding bound-state properties, though one 
should certainly not forget that this is a simplification. 
The relations in question are \cite{Herndon:1967,Gibson:1994}
\begin{eqnarray} 
{^{\,3}_\La \rm H:} \quad \tilde V_{\La N} \approx \frac{3}{4} V^s_{\La N} + \frac{1}{4} V^t_{\La N} \label{3H} \\
\label{4H0}
{^{\,4}_\La \rm He} \ (0^+): \quad \tilde V_{\La N} \approx 
\frac{1}{2} V^s_{\La N} + \frac{1}{2} V^t_{\La N} \\
\label{4H1}
{^{\,4}_\La \rm He} \ (1^+): \quad \tilde V_{\La N} \approx \frac{1}{6} V^s_{\La N} + \frac{5}{6} V^t_{\La N} \\
\label{5H}
{^{\,5}_\La \rm He:} \quad \tilde V_{\La N} \approx \frac{1}{4} V^s_{\La N} + \frac{3}{4} V^t_{\La N} 
\end{eqnarray}
From these follows the well-known fact that the 
hypertriton is dominated by the $\La N$ singlet
interaction while the ${^{\, 4}_\La \rm He} \ (1^+)$ and ${^{\,5}_\La \rm He}$
states are dominated by the triplet interaction.

Our results for the binding (separation) energies for the 
hypertriton and the $^{\,4}_\La \rm He$ hypernucleus 
are listed in Table~\ref{tab:R4new}. 
(Preliminary results for the NLO19 interaction were reported in 
\cite{Nogga:2018,Nogga:2018a} based on a different NN interaction.)  
The hypertriton binding energies for the two NLO interactions
are identical within the uncertainty caused by the
regulator dependence. The overall 
variations are of the order of $50$~keV. As noted just 
above, in this case, the binding energy is
dominated by the $\La N$ interaction in the $^1S_0$ 
(singlet) interaction, see Eq.~(\ref{3H}). That partial 
wave is less influenced by the $\La-\Si$ conversion
as can be read off from the fact that the imaginary part of the 
$\Si N$ ($I=1/2$) $^1S_0$ scattering length is 
zero for basically all considered potentials, cf. Table~\ref{tab:R0}, see also Fig.~\ref{fig:Ph1s0}. 

There is somewhat stronger variation in the predictions for 
the ${^{\,4}_\La \rm He}$ binding energies, cf. 
Table~\ref{tab:R4new}.   
However, at least for the $0^+$ state, we are reluctant 
to see a clear tendency in the results.
Recall that this state should receive contributions from 
the $^1S_0$ and $^3S_1$ $\La N$ interactions with equal 
weight, according to the simple estimate Eq.~(\ref{4H0}).
Here, the regulator dependence of the binding energy is of 
the order of $210$ and $180$~keV for NLO13 and NLO19, respectively, and, thus,
larger than the average difference between the two EFT interactions. 
Interestingly, the predictions of the two considered phenomenological $YN$ models for the $0^+$ bound
state are almost the same, despite of the large 
differences in the $\La N$-$\Si N$ transition potentials. 
Note that all considered interactions under-predict the 
experimental separation energy of the $0^+$ state. 

For the $1^+$ state of ${^{\,4}_\La \rm He}$,
the $^3S_1$ partial wave of the $\La N$ interaction
should dominate, according to Eq.~(\ref{4H1}). This
partial wave is strongly affected by the $\La-\Si$ conversion
and the effects are different for NLO13 and NLO19 as discussed in Sect.~\ref{sec:Scattering}. 
Here, we observe a more pronounced
regulator dependence of the binding energy. Specifically, 
for the NLO19 interaction, it is in the order of $300$~keV
and around $200$~keV for the NLO13 potential.   
Despite those variations, there is clearly a
trend towards larger binding energies for NLO19, i.e. for 
the interaction with a weaker $\La N$-$\Si N$ transition 
potential. This conjecture is also supported by
the result for the J\"ulich '04 potential. Here the 
transition potential in the $^3S_1-^3D_1$ partial wave 
is extremely weak and, corresponding
to that, the $1^+$ binding energy is very large. The 
prediction for the NSC97f interaction, on the other hand,
with its moderately strong transition potential matches
well with those of the chiral EFT potentials. 
Comparing with the empirical information, one can say that
the NLO19 prediction is compatible with the experiment
within the uncertainty, 
whereas the NLO13 and NSC97f interactions underestimate
the separation energy for the $1^+$ state. On the other hand, the J\"ulich '04 
potential leads to over-binding and, as a matter of fact,
to a wrong level ordering of the $0^+$ and $1^+$ states.

Similar to the ${^{\,4}_\La \rm He}$ $1^+$ state, the 
${^{\,5}_\La \rm He}$ bound state is likewise dominated 
by the $\La N$ triplet component, cf. Eq.~(\ref{5H}).
Thus, it will be interesting to see corresponding
results based on the NLO13 and NLO19 interactions~\cite{HLe}.  
The anomalously small binding energy of this state has
been notoriously difficult to describe in past 
calculations \cite{Contessi:2018}. Among other things,
a strong suppression of the $\La N$-$\Si N$ coupling
is seen as one possible explanation \cite{Gibson:1994,Nemura:2002}.
Thus, one would expect noticeable differences between 
the predictions of the two EFT interactions.   

We refrain from addressing the long-standing and still 
unsettled issue of the large charge symmetry breaking (CSB)
\cite{Gal:2016,Coon:1999,Gazda:2015,Gazda:2016} 
observed in the binding energies of 
the ${^{\, 4}_\La \rm He}$ and ${^{\, 4}_\La \rm H}$ systems  \cite{Esser:2015,Yamamoto:2015} 
here in detail. 
Indeed, there is no explicit CSB in the $\La N$ 
EFT potentials employed in the present study. Corresponding
contributions that would arise, e.g., from $\pi^0$ exchange 
in conjunction with $\La-\Si^0$ mixing \cite{Dalitz:1964}
are ignored. 
Additional CSB effects that enter into the four-body
calculations like the Coulomb interaction and the
mass difference of the $\Si^+$ and $\Si^-$ hyperons are
small \cite{Nogga:2013}. 
In Refs.~\cite{Gazda:2015,Gazda:2016}, the CSB part of the
$\La N$ interaction was constructed from the 
$\La N \to \Si N$ transition potential via an appropriate
scaling with the $\La-\Si^0$ mixing matrix element.
However, one has to be cautious in doing so. 
Specifically, one cannot turn that around and use CSB 
effects to fix the $\La N \to \Si N$ transition potential in a
quantitative way. Besides the aforementioned $\La-\Si^0$ mixing, 
there should be CSB contributions from, say, $\eta-\pi^0$ mixing 
or $\omega-\rho^0$ mixing \cite{Coon:1999} that are definitely
not proportional to the transition potential and, thus, 
demand the explicit introduction of pertinent CSB 
contact interactions in the $^1S_0$ and $^3S_1$ $\La N$ 
partial waves in the framework of EFT. 

That said, on a qualitative level there is definitely a relation between the CSB, the strength
of the $\Lambda$-$\Sigma$ conversion, and the $\Si$ component of the four-body bound-state wave
function \cite{Nogga:2013,Nogga:2002}. Therefore, we include in 
Table~\ref{tab:R4new} the probability $P_\Si$ 
to find a $\Sigma$ in the hypernuclear wave function. 
However, one should always keep in mind that this quantity is not an
observable and, thus, provides an instructive but not a real 
measure for the strength of the $\Lambda$-$\Sigma$ conversion. 
As expected, $P_\Si$ is smaller for the NLO19 interactions. There is, 
however, a visible cutoff dependence of this quantity. For NLO13 and NSC97f, 
$P_\Si$ is smaller for the $1^+$ state. This is somewhat surprising 
since Eq.~(\ref{4H1}) indicates that the triplet 
interaction should dominate and since $\Lambda$-$\Sigma$ conversion is stronger 
for the triplet in most interactions. For NLO19 and J\"ulich '04, the $1^+$ state 
has a larger $\Sigma$-probability which is more in line with naive expectations. 
As stated above, the $0^+$ separation energies are rather independent from the version 
of the chiral interaction but the $1^+$ state is more dependent this choice. For 
the $\Sigma$ probability, the dependence is exactly opposite. Therefore, it 
is clear that both properties of the interactions are not directly linked 
to each other. 

Finally, let us mention that
a new measurement by the STAR Collaboration suggests
that the ${^{\, 3}_\La \rm H}$ binding energy could
be significantly larger \cite{Adam:2019phl}. We ignore
this in the present work where the focus is on a 
comparison of our EFT interactions from 2013 and 2019.
Nonetheless, we performed some exploratory calculations
which indicate that a larger binding energy can be
indeed achieved. Moreover, the excellent description
of the $\La N$ and $\Si N$ data can be maintained, 
by an appropriate re-adjustment of the potential
strengths in the $\La N$ $^1S_0$ and $^3S_1$ partial
waves - though at the expense of giving up the
strict SU(3) constraints on the ($S$-wave) LECs
between the $\La N$ and $\Si N$ channels. 
Details will be reported elsewhere \cite{NNN}. 

\section{Discussion}
\label{sec:Discussion}

The $\La$-$\Si$ conversion and its impact
on hyperonic few- and many-body systems has been 
discussed in numerous works in the past
\cite{Nogami:1970,Bodmer:1971,Dabrowski:1973,Carlson:1991,Gibson:1994,Gibson:1988,Afnan:1989,Afnan:1990,Hiyama:2001,Nemura:2002}. 
However, in basically all studies so far simplified potential models for the 
$YN$ interactions have been employed and usually only the
extreme scenarios of ``coupled or not-coupled'' were compared.
The present study is on a much more subtle level.
First, the full complexity of the $YN$ interaction
is taken into account. Second, the coupling
of the $\La N$ and $\Si N$ is always considered
and a simultaneous description of the available
low-energy $\La p$ and $\Si N$ data is achieved
by both $YN$ potentials compared in this work.  

Nevertheless, the effects due to the $\La$-$\Si$ conversion 
revealed by the present study are qualitatively rather similar 
to those reported in earlier calculations.
This is true for three- and four-body systems \cite{Gibson:1994,Gibson:1988,Afnan:1989}  
but also for the in-medium properties of the $\La$ 
hyperon \cite{Nogami:1970,Bodmer:1971,Dabrowski:1973}. 
Perhaps surprising at first sight,  it is an indication that most interactions used 
in the former works captured reasonably well the bulk properties of the $YN$ interaction.

There is one aspect, however, that has not been really 
in the focus of past discussions and, thus, we
want to elaborate on it in more detail. It concerns
the situation embodied by the two EFT interactions: 
These yield practically identical results for $\La p$ 
as well as $\Si N$ observables in the low-energy region,
as demonstrated in Sect.~\ref{sec:Scattering},
but are characterized by a noticeably different 
strength of the $\La N \to \Si N$ transition potential.
One might think that additional and/or more accurate 
scattering data could facilitate a discrimination.
But this is unlikely, because one has to realize that the
transition potential itself is not an observable quantity. 
The situation is analogous to that of the deuteron. 
It is well-known that its $D$-state probability 
is not a measurable quantity \cite{Friar:1979}. Yet it cannot 
be zero (because of the quadrupole moment
of the deuteron) and not too large either. Similarly,
the measured $\Si^- p \to \La n$ (and $\La p \to \Si^0 p$)
cross section requires a non-zero transition potential,
but it fixes its actual strength only within certain limits. 

In few- and many-body calculations involving hyperons, 
differences in the elementary $\La N \to \Si N$ 
transition potential are to be balanced by corresponding 
three-body forces (3BFs). In chiral EFT, the latter appear 
naturally and automatically in a consistent implementation of 
the framework \cite{Epelbaum:2006,Epelbaum:2008,Hammer:2013}. 
In the power counting followed in Ref. \cite{Haidenbauer:2013}
and in the present work, such 3BFs arise first at next-to-next-to-leading order (N$^2$LO)
in the chiral expansion \cite{Epelbaum:2006,Petschauer:2015BF}.
For the specific case of the $\La$-$\Si$ conversion, the 
necessity for 3BFs is illustrated in a pedagogical 
way by the similarity renormalization group (SRG) transformation,
a tool that is nowadays commonly applied in studies of few-nucleon systems but also of 
hypernuclei~\cite{Bogner:2009bt,HLe,Wirth:2014,Wirth:2016,Wirth:2018,Wirth:2019}. 
It amounts to a prediagonalization of the Hamiltonian in  momentum space
in order to improve the convergence of calculations using various many-body methods. One
specific feature of this diagonalization is the occurrence 
of so-called induced three- and higher many-body forces of moderate size. 
In applications to hypernuclei, such a prediagonalization
also involves a decoupling of the $\La N$ and $\Si N$ 
systems, i.e. leads to a strong reduction of the 
$\La N \to \Si N$ transition potential in the Hamiltonian. 
In this case, induced $YNN$ three-body forces appear,
however, they have a more sizable effect as discussed in 
detail in Ref.~\cite{Wirth:2016}. This clearly demonstrates 
that in few- and many-body applications the actual strength 
of the $\La N \to \Si N$ transition potential 
is correlated with and has to supplemented by that of 
corresponding ($\La NN$, $\Si NN$) three-body forces.    

\begin{figure}
\begin{center}
\includegraphics[height=180mm]{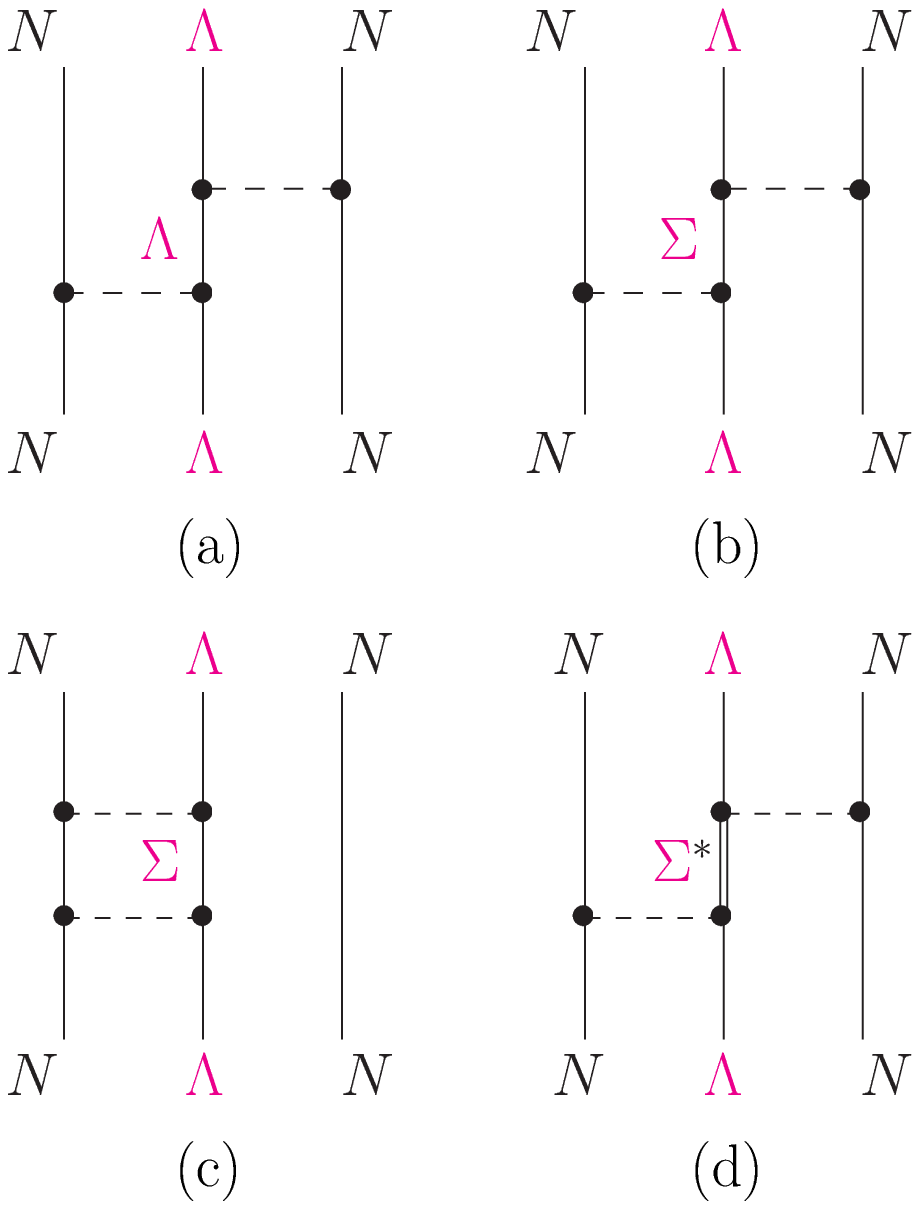}
\vskip -9cm 
\caption{Illustration of three-body dynamics:
(a) standard contribution in the $\La NN$ three-body
equations. 
(b) effective three-body force arising from the 
$\La N$-$\Si N$ coupling.
(c) $\La N$-$\Si N$ transition in the presence
of a spectator, leading to a dispersive effect. 
(d) three-body force due to the excitation of the
$\Si^*$(1385) baryon. 
}
\label{fig:3BF}
\end{center}
\end{figure}

Let us further expand on the role of three-body 
forces in few-body system. To begin with we want
to remind the reader that 3BFs are strongly scheme dependent, 
as discussed extensively in, e.g., Ref.~\cite{Hammer:2013}. 
Specifically, the actual physics represented by a 3BF depends 
crucially on the degrees of freedom taken into account 
in the effective field theory and/or in the specific calculation. 
For example, in the SU(3) chiral EFT applied by us, the $\La$ and 
$\Si$ hyperons are treated on equal footing. This is also 
done in the three- and four-body calculations within the 
conventional Faddeev-Yakubovsky approach presented in 
Sect.~3.3. Then contributions represented schematically
by the diagrams (a)-(c) in Fig.~\ref{fig:3BF} are all 
automatically included by solving
the corresponding Eqs.~(\ref{3BE}) and (\ref{4BE}). 
As discussed thoroughly in Ref.~\cite{Miyagawa:1993},
the inclusion of the $\Si$ 
leads to two types of effects in the three-body 
dynamics. One is the so-called dispersive effect. It 
arises when the $\La N$ interaction takes place in the 
presence of one or two spectator nucleons, cf. 
Fig.~\ref{fig:3BF}~(c). Then the contribution of, 
say, the transition $\La N \to \Si N \to \La N$ to the 
effective two-body potential is reduced as compared to 
the situation in free space because the propagator
includes now the kinetic energy of the spectator
nucleons and, as a consequence, the effective 
interaction is less attractive \cite{Miyagawa:1993,Gibson:1994}. 
At the same time, the equal treatment of the
$\La$ and $\Si$ hyperons in the two- and few-body 
equations generates contributions of the form shown 
in Fig.~\ref{fig:3BF}~(b). In SU(3) chiral EFT, this 
contribution is not a 3BF \cite{Petschauer:2015BF}
but a result of two-body dynamics. 
The corresponding contributions can be attractive and 
then they can compensate or even exceed the dispersive 
effects. Note that a smaller (larger)
$\La N \to \Si N$ transition potential leads to smaller (larger)
dispersive effects but at the same time also 
to smaller (larger) ``3BFs'', so that the net result 
might be not too sensitive to the actual strength of the
transition potential, provided that a consistent and 
complete calculation has been performed as in the present 
study. Of course, in general, the properties of the  3BF-type contributions
generated in this way depend crucially on the considered 
state and hypernucleus so that there will be a delicate 
and distinct interplay between the two three-body effects. 
The diagram in Fig.~\ref{fig:3BF}~(d) is not generated 
by the dynamical equations since decuplet baryons are
not explicitly included. Thus, it constitutes a 
proper contribution to the 3BF in SU(3) chiral EFT
\cite{Petschauer:2015BF,Petschauer:2016BF}.  

The situation is different for pionless EFT which has been 
likewise employed in studies of the properties of the 
hypertriton and of other light hypernuclei 
\cite{Hammer:2002,Ando:2015,Contessi:2018,Hildenbrand:2019}. 
In that framework, only the $\La$ and the nucleons are
kept as active degrees of freedom while pions but also 
the $\Si$ are ``frozen out''. As a consequence, proper 
3BFs appear already at LO in this approach. And these 3BFs
represent effectively the dynamics illustrated in 
Fig.~\ref{fig:3BF} (b), (c) and (d), among other things.
Recall that the virtual elimination of the $\Si$ degrees of 
freedom is also one of the reasons for the induced 3BFs
in the discussed SRG transformation \cite{Wirth:2016}.  

Studies of the nuclear matter properties are usually based 
on the $G$-matrix calculated from the Bethe-Gold\-stone equation,
as it is done here. Then only the dispersive effect is
taken into account and, thus, a stronger $\La N\to \Si N$
potential leads unavoidably to a less attractive
$\La$ nuclear potential. It is the prime reason 
why we see a sizable difference in the nuclear 
matter results for the NLO13 and NLO19 potentials, 
cf. Sect.~\ref{sec:Matter}. But it is also the
main reason for the differences in nuclear matter calculations
observed for phenomenological $YN$ potentials. 
A proper way to deal with this would be to solve the corresponding Bethe-Faddeev
equation \cite{Bethe:1965} where three-body correlations 
are taken into account consistently. It is, however, 
technically rather challenging and therefore commonly
avoided. 

Should one give preference to either the NLO13 or the
NLO19 interaction? In our opinion there are no stringent 
reasons that would make one superior over the other. 
That said, a $YN$ potential where SU(3) symmetry is 
fulfilled by the NLO LECs in combination with the $NN$
interaction and which is, therefore, more 
in line with the underlying power counting, is 
certainly more favorable from a fundamental point of view. 
Note, however, that the symmetry is anyway broken by 
the corresponding NLO contributions from two-meson exchange
\cite{Haidenbauer:2013}. 
Certainly a positive aspect in favor of the new fit 
is that the corresponding LECs are somewhat smaller and, 
therefore, more in line with the requirement of 
natural size \cite{Epelbaum:2006}.

Another aspect is, whether it would be sensible to ``optimize''  
the $YN$ potential so that the 3BFs become small. 
As argued in Ref.~\cite{Hammer:2013}, such a 
strategy is doomed to fail on the level of accuracy
of the last order of the EFT where 3BFs do not contribute. 
For a 3BF that only depends on one adjustable parameter, 
it is obviously advantageous to chose the $YN$ interaction such 
that this parameter is minimal. This 
will simultaneously minimize the effect of the 3BF 
in all observables. 
Once several parameters are involved, as is certainly 
the case for the chiral $YNN$ interaction at N$^2$LO, 
the optimal set of parameters 
will depend on the observable chosen and will not be universal.
A good example of this aspect shown in this work is the 
observation that we can improve the description of 
the $1^+$ state for specific choices of the cutoff or by choosing 
NLO19. This, however, will not improve the description of the 
$0^+$ state. 
State-of-the-art calculations of few-nucleon systems 
based on chiral $NN$ potentials commonly include 3BFs. 
The arising additional LECs in the 3BF are fixed by 
considering few- or many body observables, for example 
the triton binding energy or the minimum of the 
differential $pd$ cross section \cite{Epelbaum:2019}. 
Their actual values depend on the specific features 
of the employed EFT and, in particular, on the adopted 
regularization scheme \cite{Hammer:2013}.  
With regard to few-body systems involving hyperons, the 
LECs corresponding to $\La NN$ (or $\Si NN$) 
forces will be necessarily interrelated with the 
strength of the $\La N \to \Si N$ transition potential. 
Since such 3BFs arise only at N$^2$LO in the power 
counting, as mentioned above, we do not 
consider them in the present work which is at the NLO level.
Anyway, given the present experimental situation
it remains unclear in how far the various LECs that 
arise in the lowest-order $\La NN$ 3BF \cite{Petschauer:2015BF} 
could be fixed by considering few-body observables such as 
the ${^{\, 3}_\La \rm H}$ and/or 
${^{\, 4}_\La \rm H}$ (${^{\, 4}_\La \rm He}$) binding
energies.
One possible solution would be the explicit inclusion of
decuplet baryons in the EFT as discussed in Ref.~\cite{Petschauer:2016BF}.
Assuming that the $YNN$ 3BF can be saturated by the
excitation of decuplet baryons reduces the number of pertinent
LECs considerably. As a byproduct, this framework
would also promote the corresponding
contributions to NLO \cite{Epelbaum:2008} which is consistent with 
the assumption that these contributiuons are the dominate 3BFs 
to be expected in chiral EFT.  

Either way, including 3BFs into our 
codes for solving the Faddeev-Yakubovsky equations for 
the $YNN$ and $YNNN$ systems is technically rather demanding.
It requires considerable additional work which we postpone 
to the future.   Thus, at present, we cannot give reliable estimates for the 
size of 3BFs. However, since the effect of the $\La$-$\Si$ conversion 
is explicitly included in our few-body calculation, 
we expect only moderate contributions from such 3BFs 
for light hypernuclei and, specifically for the hypertriton.
This is in contrast with the aforementioned studies within 
pionless EFT \cite{Hammer:2002,Contessi:2018} 
or with phenomenological approaches \cite{Lonardoni:2013} where 
an effective $\La N$ interaction is employed and the coupling 
to $\Si N$ is not taken into account. Here the effect of 3BFs 
is significant and without including them explicitly, no 
realistic results can be achieved, as testified by past 
calculations. 
It is argued in Refs.~\cite{Nogga:2013,Hammer:2013} 
that the dependence of the predictions on the regulator 
should provide a lower bound for the magnitude of
the contributions from three- and higher-body forces. 
Based on that measure, one expects a rather small influence
in case of the hypertriton. This is in line with other arguments
that consider the fact that the bound state is very shallow
and, accordingly, the $\La$ is on average far from the 
two nucleons \cite{Miyagawa:1995}. Then the likelihood that
all three particles are close to each other and feel a
3BF is very small. 
For the ${^{\, 4}_\La \rm He}$ system, the cutoff dependence
of the separation energies is larger and, thus, one would expect 
larger effects from 3BFs here, specifically for the 
$1^+$ state. 

\begin{table}
  \caption{Comparison of the separation energies $E_\Lambda$ of $^3_\Lambda {\rm H}$, $^4_\Lambda {\rm He(0^+)}$
    and $^4_\Lambda {\rm He(1^+)}$ including and excluding 
    explicit $\Sigma$s for different $YN$ interactions. 
    See text for more details. 
    For the chiral $YN$ forces, the SMS 
    $NN$ interaction \cite{Reinert:2017} at order N$^4$LO+ 
    with cutoff of 450~MeV has been used. For J\"ulich'04 and NSC97f, the CD-Bonn interaction \cite{CDBonn}
    has been employed.     
    Energies are given in MeV. 
    }
    \label{tab:R4woSigma}
    \centering
    \renewcommand{\arraystretch}{1.4}
\begin{tabular}{|r|ccc|}
\hline
      YN interaction & $^3_\Lambda {\rm H}$  & $^4_\Lambda {\rm He(0^+)}$  & $^4_\Lambda {\rm He(1^+)}$ \\
\hline
 NLO13(650) w/   $\Sigma$ &   0.087   &    1.490 &   0.615  \\
 NLO13(650) w/o  $\Sigma$ &   0.095   &    1.155 &   0.568  \\
 \hline
 NLO19(650) w/   $\Sigma$ &   0.095   &    1.530 &   0.916 \\
 NLO19(650) w/o  $\Sigma$ &   0.100   &    1.300 &   0.735 \\
 \hline 
 J\"ulich'04 w/  $\Sigma$ &   0.046   &    1.704 &   2.312 \\
 J\"ulich'04 w/o $\Sigma$ &   0.162   &    2.397 &   2.319 \\     
 \hline 
 NSC97f w/ $\Sigma$       &   0.099   &    1.832 &   0.575 \\
 NSC97f w/o $\Sigma$      &   0.062   &    1.303 &   0.679  \\
 \hline 
 \end{tabular}  
 \renewcommand{\arraystretch}{1.0}
\end{table}

A rough indication for the magnitude of possible effects 
of 3BFs can be obtained by switching off the 
$\Si$ in the three- and four-body systems as discussed in 
Refs.~\cite{Nogga:2002,Nogga:2013}. 
Corresponding results are summarized in Table~\ref{tab:R4woSigma}. 
Clearly, this procedure provides primarily a measure for the 
effective 3BFs coming from the $\Si$ excitation, cf. 
Fig.~\ref{fig:3BF} (b). 
But one might speculate that the magnitude of an 
actual 3BF represented, e.g., by the excitation of 
the $\Si^*$(1385) \cite{Petschauer:2016BF} see Fig.~\ref{fig:3BF}~(c), should 
be smaller given that the $\Si^*$ mass 
is significantly larger and that the power counting expects first contributions 
at a higher order. 
The actual change in the $^3_\Lambda$H separation  
energy for the hypertriton amounts to less than 
$10$~keV for the NLO interactions when the 
$\Si$ component is switched off. There is an increase 
in the binding which means that the 
effective 3BFs coming from the $\Si$ excitation are
overall repulsive. Interestingly, the opposite is the case
for the NSC97f potential, and also for other Nijmegen
$YN$ interactions considered in the past 
\cite{Nogga:2013,Nogga:2002}. Obviously, there is a 
delicate interplay reflecting the actual strength of 
the $\La$-$\Si$ conversion as well as its realization 
in the $^1S_0$ and $^3S_1$ partial waves. 
In the four-body system, there is a reduction of the binding 
energy by around 
$340$ ($230$)~keV for the $0^+$ state and by 
$150$ ($180$)~keV for $1^+$, for NLO13 (NLO19), when 
the $\Si N$ component is switched off in the few-body
calculations.  For results with the NLO13 interaction with 
other cutoffs, see Ref.~\cite{Nogga:2013}. 
Also for $^4_\Lambda {\rm He}$, the trend exhibited by 
the phenomenological potentials differs in part. 
Nonetheless, at least for the chiral interactions, 
the variations in the separation energies when the $\Sigma$ component 
is removed is even slightly smaller than the cutoff dependence, 
discussed above. Since these variations provide a measure for 
the diagram of Fig.~\ref{fig:3BF} (b), the results support 
that 3BFs in our approach \cite{Petschauer:2015BF} are likely smaller than the 
uncertainty at order NLO. 

Finally, note that, for nuclear matter calculations, one possibility 
to circumvent the computational challenges of many-body 
equations consists in the use of density-de\-pend\-ent 
effective $\La N$ (and $\Si N$) interactions that can be
derived from chiral three-body forces
\cite{Petschauer:2015BF}. 
Assuming furthermore that the 3BFs are dominated by 
the excitation of decuplet baryons (decuplet saturation), 
the number of independent LECs in the three-baryon 
interactions can be considerably reduced. 
A first application of that formalism in studies
of the in-medium properties of the $\La$ has been
reported in Ref.~\cite{Haidenbauer:2017}.
In this context, let us mention that adding a
density-dependent effective $\La N$ force to the
NLO19 interaction, with the strength considered in the
aforementioned reference, would bring the 
single-particle potential $U_\La$ for NLO13 and NLO19
roughly in agreement with each other, up to the highest
considered Fermi momentum of $k_F = 1.7$~fm$^{-1}$,
corresponding to a density of twice the nuclear
matter saturation density. 

\section{Conclusions}
\label{sec:Conclusions}

In the present work we have investigated the $\Lambda N$ 
and $\Sigma N$ interactions at next-to-leading order in 
SU(3) chiral effective field theory. In particular, we
have explored different options for the low-energy constants 
that determine the strength of the contact interactions.
One $YN$ interaction considered is the initial NLO potential
published in 2013 \cite{Haidenbauer:2013}. The other potential
has been established in the present paper. It is guided by
the objective to reduce the number of LECs that need to be
fixed in a fit to the $\Lambda N$ and $\Sigma N$ data
by inferring some of them from the $NN$ sector via the
underlying SU(3) symmetry. Correlations between
the LO and NLO LECs of the $S$-waves had been observed 
already in our initial $YN$ study \cite{Haidenbauer:2013} and 
indicated that a unique determination of them by considering
the existing $\Lambda N$ and $\Sigma N$ data alone 
is not possible. 

As demonstrated in the present work, the two variants considered
yield equivalent results for $\Lambda N$ and $\Sigma N$ scattering
observables. However, they differ in the strength of the $\Lambda N \to \Sigma N$
transition potential and that becomes manifest in 
applications to few- and many-body systems. 
The influence of this difference on predictions for 
light hypernuclei and for the properties
of the $\Lambda$ and $\Sigma$ hyperons in nuclear 
matter has been shown and discussed in detail. 
It turned out that the effect of the variation 
in the strength of the $\Lambda N$-$\Sigma N$ coupling
($\La -\Si$ conversion) is moderate for the considered hypernuclei
but sizable in case of the matter properties. 

Since the $YN$ scattering data alone cannot fully constrain 
the $\Lambda N$-$\Sigma N$ transition potential, 
arguably as a matter of principle, 
consistent three-body forces are needed to compensate
for the differences in few- and many-body systems. 
Such 3BFs arise only at N$^2$LO in the power 
counting that we follow, and therefore, we did not 
consider them in the present work which is at the 
NLO level. However, we speculate that the effect of
such 3BFs should be fairly small, at least for
light hypernuclei, once the $\La -\Si$ conversion
is taken into account consistently in the corresponding
calculations. In such a case, important aspects of
three-body dynamics such as dispersive effects but
also effective three-body forces that arise from the
coupling of $\Lambda N$ to $\Sigma N$ are taken
into account rigorously. 

In this work, the influence of the $\Lambda-\Sigma$ 
conversion strength on light hypernuclei and nuclear matter
has been investigated. For further insight, but also for 
addressing other aspects, more and/or more accurate data 
are essential. 
A new measurement of the hypertriton bound state has 
been presented which points to a noticeably larger 
binding energy \cite{Adam:2019phl}. 
Measurements of $^{\,4}_\La \rm H$ and
$^{\,4}_\La \rm He$ with improved accuracy in order 
to determine the amount of charge-symmetry breaking 
more precisely have been performed
\cite{Esser:2015,Yamamoto:2015} or are on the way
\cite{JPARC}.  
There are also attempts to shed more light on the 
elementary $YN$ interaction directly via studies
of the $\La p$ correlation function measured in 
heavy-ion collisions or high-energy $pp$ collisions
\cite{Adams:2006,Anticic:2011,Acharya:2019}. Moreover,
there are plans for a future measurement of $\Si^-p$ 
scattering at J-PARC \cite{Miwa:2019}. 
Depending on the outcome of those experiments one might
have to readjust the $YN$ interaction. 
In particular, this concerns the relative strength of 
the $\La p$ interaction in the $^1S_0$ and $^3S_1$ 
channels. Efforts at the COSY accelerator
in J\"ulich to determine the strength of the spin-triplet 
$\La p$ interaction from the final-state interaction 
in the reaction $pp\to K^+\Lambda p$ \cite{Hauenstein:2017}
suffered from low statistics and, unfortunately, could 
not provide the desired and urgently needed stringent
constraint. Given the lack of appropriate spin-dependent
observables, it is fixed in our studies by considering the 
hypertriton separation energy  
\cite{Haidenbauer:2013,Haidenbauer:2005,Polinder:2006}.   
A larger hypertriton separation energy would certainly require
a more attractive $^1S_0$ $\La p$ interaction. 
That, in turn, would influence the predictions of an 
appropriately modified chiral $Y N$ interaction for 
the $^{\,4}_\La \rm H$ and $^{\,4}_\La \rm He$ states. Work 
in this direction is already in progress \cite{NNN}. 

\begin{acknowledgement}
This work is supported in part by the DFG and the NSFC through
funds provided to the Sino-German CRC 110 ``Symmetries and
the Emergence of Structure in QCD'' (DFG grant. no. TRR~110)
and the VolkswagenStiftung (grant no. 93562).
The work of UGM was supported in part by The Chinese Academy
of Sciences (CAS) President's International Fellowship Initiative (PIFI)
(grant no.~2018DM0034). The numerical calculations were performed on JURECA
and the JURECA-Booster of the J\"ulich Supercomputing Centre, J\"ulich, Germany.
\end{acknowledgement}


\section*{Appendix: Contribution of contact terms}
\label{sec:Appendix}

A detailed description of the derivation of the hyperon-nucleon interaction within
SU(3) chiral EFT up to NLO, based on the modified Weinberg counting \cite{Epelbaum:2005}, 
has been given in Ref.~\cite{Haidenbauer:2013}. 
Specifically, in this work, explicit ready-to-use expressions for the potentials 
in the $\La N \to \La N$ and $\Si N \to \Si N$ channels, and the $\La N \to \Si N$ 
transition can be found. Since the contributions from one- and two-meson exchanges 
of the Goldstone bosons $\pi$, $\eta$, and $K$, included in the present work, are
identical to those in \cite{Haidenbauer:2013}, we refrain from reproducing
the pertinent formulae and tables here. We do, however, provide the 
expressions for the contact terms and the low-energy constants (LECs) 
associated with them because the latter are the quantities that have 
been re-adjusted for the potential presented in this work. 
In addition the relations between the LECs for the various $YN \to Y'N$ 
transition potentials, that follow from the assumed SU(3) symmetry, are given. 

\begin{table*}[ht]
\caption{SU(3) relations for the various contact potentials in the isospin basis.
$C^{27}_{\xi}$ etc. refers to the corresponding irreducible SU(3) representation
for a particular partial wave ${\xi}$. The actual potential still needs to be
multiplied by pertinent powers of the momenta $p$ and $p'$.
The same relations hold for $\tilde C^{27}_{\xi}$ etc.
}
\label{tab:SU3}
\vskip 0.1cm
\renewcommand{\arraystretch}{1.2}
\centering
\begin{tabular}{|l|c|c|l|l|}
\hline
&Channel &I &\multicolumn{2}{|c|}{$V({\xi})$} \\
\hline
&        &  &$\xi= \, ^1S_0  $
& $\xi = \, ^3S_1, \, ^3S_1$-$^3D_1  $  \\
\hline
${S=\phantom{-}0}$&$NN\rightarrow NN$ &$0$ & \ \ -- & $C^{10^*}_{\xi}$ \\
                       &$NN\rightarrow NN$ &$1$ & $C^{27}_{\xi}$ & \ \ -- \\
\hline
${S=-1}$&$\La N \rightarrow \La N$ &$\frac{1}{2}$ &$\frac{1}{10}\left(9C^{27}_{\xi}+C^{8_s}_{\xi}\right)$
& $\frac{1}{2}\left(C^{8_a}_{\xi}+C^{10^*}_{\xi}\right)$ \\
&$\La N \rightarrow \Si N$ &$\frac{1}{2}$        &$\frac{3}{10}\left(-C^{27}_{\xi}+C^{8_s}_{\xi}\right)$
& $\frac{1}{2}\left(-C^{8_a}_{\xi}+C^{10^*}_{\xi}\right)$ \\
&$\Si N \rightarrow \Si N$  &$\frac{1}{2}$        &$\frac{1}{10}\left(C^{27}_{\xi}+9C^{8_s}_{\xi}\right)$
& $\frac{1}{2}\left(C^{8_a}_{\xi}+C^{10^*}_{\xi}\right)$ \\
&$\Si N \rightarrow \Si N$  &$\frac{3}{2}$        &$C^{27}_{\xi}$ & $C^{10}_{\xi}$ \\
\hline
\end{tabular}
\renewcommand{\arraystretch}{1.0}
\end{table*}

The spin dependence of the potentials due to leading order contact terms is 
given by \cite{Haidenbauer:2013} 
\begin{eqnarray}
V^{(0)}_{YN\to Y'N} &=& C_{S} + C_{T}\,
\mbox{\boldmath $\sigma$}_1\cdot\mbox{\boldmath $\sigma$}_2\,,
\end{eqnarray}
where the parameters $C_{S}$ and $C_{T}$ are the aforementioned LECs, 
which depend on the considered $YN$ baryon-baryon channel 
and which need to be determined in a fit to data.
At next-to-leading order, the spin- and momentum-depen\-dence of the contact terms reads
\begin{eqnarray}
V^{(2)}_{YN\to Y'N} &=& C_1 {\bf q}^{\,2}+ C_2 {\bf k}^{\,2} + (C_3 {\bf q}^{\,2}+ C_4 {\bf k}^{\,2})
\,\mbox{\boldmath $\sigma$}_1\cdot\mbox{\boldmath $\sigma$}_2 \nonumber \\
&+& \frac{i}{2} C_5 (\mbox{\boldmath $\sigma$}_1+\mbox{\boldmath $\sigma$}_2)\cdot ({\bf q} \times {\bf k}) 
+ C_6 ({\bf q} \cdot \mbox{\boldmath $\sigma$}_1) ({\bf q} \cdot \mbox{\boldmath $\sigma$}_2) \nonumber \\
&+& C_7 ({\bf k} \cdot \mbox{\boldmath $\sigma$}_1) ({\bf k} \cdot \mbox{\boldmath $\sigma$}_2)
+ \frac{i}{2} C_8 (\mbox{\boldmath $\sigma$}_1-\mbox{\boldmath $\sigma$}_2)\cdot ({\bf q} \times {\bf k}) \ , 
\nonumber \\
\end{eqnarray}
where  $C_i$ ($i=1,\dots,8$) are additional LECs.
The transferred and average momenta, ${\bf q}$ and ${\bf k}$, are defined in terms of
the final and initial center-of-mass momenta of the baryons, ${\bf p}'$ and ${\bf p}$, as
${\bf q}={\bf p}'-{\bf p}$ and ${\bf k}=({\bf p}'+{\bf p})/2$.
When performing a partial-wave projection, these terms contribute to the two $S$--wave
($^1S_0$, $^3S_1$) potentials, the four $P$--wave
($^1P_1$, $^3P_0$, $^3P_1$, $^3P_2$) potentials, and the $^3S_1$-$^3D_1$ and $^1P_1$-$^3P_1$
transition potentials in the way described in Sec. 2.1 of Ref.~\cite{Haidenbauer:2013}.
For the $^1S_0$ and $^3S_1$-$^3D_1$ partial waves relevant here, these can be cast in
the form
\begin{eqnarray}
\label{VC0}
V(^1S_0) &=& \tilde{C}_{^1S_0} + {C}_{^1S_0} ({p}^2+{p}'^2)~, \label{C1S0}\\
V(^3S_1) &=& \tilde{C}_{^3S_1} + {C}_{^3S_1} ({p}^2+{p}'^2)~, \label{C3S1} \\
V(^3D_1 -\, ^3S_1) &=& {C}_{^3S_1 -\, ^3D_1}\, {p'}^2~,\\
V(^3S_1 -\, ^3D_1) &=& {C}_{^3S_1 -\, ^3D_1}\, {p}^2~, 
\end{eqnarray}
with $p = |{\bf p}\,|$ and ${p}' = |{\bf p}\,'|$.

The SU(3) structure is summarized in Table~\ref{tab:SU3}. Here the LECs are 
expressed in terms of the irreducible representations resulting from the decomposition 
of the tensor product relevant for the scattering of two octet baryons:
$8$ $\otimes$ $8$ = $1$ $\oplus$ $8_a$ $\oplus$ $8_s$ $\oplus$ $10^*$ $\oplus$ $10$ $\oplus$ $27$
(for details see Refs.~\cite{Swart:1963,Dover:1991}). 
From that table, one can immediately read off the potential for a specific
$YN\to Y'N$ transition and a specific partial wave. It is simply a combination 
of the SU(3) structure and the spin-momentum structure and reads, 
for example, for the $^1S_0$ partial wave of the
$\La N \to \La N$ channel:
\begin{eqnarray}
&&V_{\La N \to \La N}(^1S_0) = \nonumber \\
&&\frac1{10}\left[9\tilde C^{27}_{^1S_0} + \tilde C^{8_s}_{^1S_0}
+ (9C^{27}_{^1S_0} + C^{8_s}_{^1S_0})(p^2+p'^2)\right]. 
\end{eqnarray}

In the fitting procedure, the ``standard'' set of 36 $YN$ data points \cite{Polinder:2006} 
has been included, which consists of low-energy total cross sections for the 
reactions:
$\Lambda p \to \Lambda p$
from Ref.~\cite{Sec68} (6 data points) and Ref.~\cite{Ale68} (6 data points),
$\Sigma^- p \to \Lambda n$
\cite{Eng66} (6 data points),
$\Sigma^- p \to \Sigma^0 n$
\cite{Eng66} (6 data points),
$\Sigma^- p \to \Sigma^- p$
\cite{Eis71} (7 data points),
$\Sigma^+ p \to \Sigma^+ p$
\cite{Eis71} (4 data points),
and the inelastic capture ratio at rest \cite{Hep68,Ste70}.
Besides these $YN$ data, the empirical $\Lambda$ separation energy of the hypertriton
$^3_\La \rm H$ of 130~keV \cite{Davis:1992dt}
has been used as a further constraint.
Without the latter it would not be possible to fix the relative strength of the
spin-singlet and spin-triplet $S$-wave contributions to the $\Lambda p$ interaction.
The actual values of the employed LECs are summarized in Table~\ref{tab:LECs}. 
The LECs for the $P$-waves are all taken over from Ref.~\cite{Haidenbauer:2013}.
Their values can be found in Table~4 of that work. 

\begin{table*}
\caption{Contact terms for the $^1S_0$ and $^3S_1$-$^3D_1$ $YN$ 
partial waves for various cutoffs $\Lambda$. 
The values of the $\tilde C$'s are in $10^4$ ${\rm GeV}^{-2}$ 
the ones of the $C$'s in $10^4$ ${\rm GeV}^{-4}$. 
The values of $\Lambda$ are in MeV.
}
\renewcommand{\arraystretch}{1.4}
\label{tab:LECs}
\vspace{0.2cm}
\centering
\begin{tabular}{|c|c|rrrr|}
\hline
\multicolumn{2}{|c|}{$\Lambda$} & $500$ & $550$& $600$& $650$ \\
\hline
$^1S_0$
&$\tilde C^{27}_{^1S_0}$    &$-0.0062$ &$0.0922$   &$0.2564$  &$0.5375$  \\
&$C^{27}_{^1S_0}$           &$2.313$   &$2.326$    &$2.326$   &$2.328$  \\
&$\tilde C^{8_s}_{^1S_0}$   &$0.1970$  &$0.1930$   &$0.1742$  &$0.1670$  \\
&$C^{8_s}_{^1S_0}$          &$-0.2000$ &$-0.2060$  &$-0.0816$ &$-0.0500$  \\
\hline
$^3S_1$-$^3D_1$
&$\tilde C^{10^*}_{^3S_1}$  &$-0.0987$  &$-0.0476$  &$0.2198$   &$0.6688$   \\
&$C^{10^*}_{^3S_1}$         &$0.2977$   &$0.3139$   &$0.5109$   &$0.4899$   \\
&$\tilde C^{10}_{^3S_1}$    &$0.3322$   &$0.4390$   &$0.6672$   &$0.8961$   \\
&$C^{10}_{^3S_1}$           &$0.6799$   &$0.6910$   &$0.4681$   &$0.4200$   \\
&$\tilde C^{8_a}_{^3S_1}$   &$0.0236$   &$0.0393$   &$0.0279$   &$-0.0021$  \\ 
&$C^{8_a}_{^3S_1}$          &$0.3955$   &$0.3745$   &$0.4496$   &$0.6589$   \\
&$C^{10^*}_{^3S_1-\,^3D_1}$ &$-0.2406$  &$-0.2595$  &$-0.2422$  &$-0.1913$   \\
&$C^{10}_{^3S_1-\,^3D_1}$   &$-0.3000$  &$-0.1115$  &$-0.3800$  &$-0.3638$  \\ 
&$C^{8_a}_{^3S_1-\,^3D_1}$  &$0.1728$   &$-0.0136$  &$-0.0348$  &$-0.0437$  \\ 
\hline
\end{tabular}
\renewcommand{\arraystretch}{1.0}
\end{table*}


\end{document}